\documentclass[aps,prb,twocolumn,superscriptaddress,nofootinbib,longbibliography]{revtex4-1}

\usepackage{amsmath,amssymb}
\usepackage{amsfonts}
\usepackage{slashed}	
\usepackage{feynmp-auto} 
\usepackage{physics}

\usepackage[export]{adjustbox}

\usepackage{graphicx}
\usepackage{fancybox,color}

\usepackage{float}

\usepackage[utf8]{inputenc} 
\usepackage[T1]{fontenc} 

\usepackage[font=footnotesize,labelfont=bf, justification=centerlast]{caption}

\usepackage{xspace}

\usepackage[english]{babel}
\usepackage{lmodern}

\usepackage{amsthm}
\theoremstyle{remark}
\newtheorem{example}{Example}

\def\ra{\rightarrow}
\def\tra{$\rightarrow$ }

\def\1{\mathbf{1}}

\def\L{{\cal L}}
\renewcommand{\ell}{L}

\def\W{{\cal W}}
\def\>{\rangle}
\def\<{\langle}

\def\h{}

\def\s{{}^\dagger}
\def\eps{\varepsilon}
\def\la{\lambda}
\def\si{\sigma}
\def\ga{\gamma}
\def\Ga{\Gamma}

\def\su{\uparrow}
\def\sd{\downarrow}
\def\Eb{E}

\def\App#1{} 

\begin{document}

\title{Relaxation dynamics in a Hubbard dimer coupled to fermionic baths: phenomenological description and its microscopic foundation}

\author{Eric Kleinherbers}
\email{eric.kleinherbers@uni-due.de}
\affiliation{Fakult\"at f\"ur Physik, Universit\"at Duisburg-Essen, Lotharstra{\ss}e 1, 47057 Duisburg, Germany,}
\affiliation{CENIDE, Universit\"at Duisburg-Essen, Carl-Benz-Straße 199, 47057 Duisburg, Germany,}
\author{Nikodem Szpak}
\email{nikodem.szpak@uni-due.de}
\affiliation{Fakult\"at f\"ur Physik, Universit\"at Duisburg-Essen, Lotharstra{\ss}e 1, 47057 Duisburg, Germany,}
\author{J\"urgen K\"onig}
\affiliation{Fakult\"at f\"ur Physik, Universit\"at Duisburg-Essen, Lotharstra{\ss}e 1, 47057 Duisburg, Germany,}
\affiliation{CENIDE, Universit\"at Duisburg-Essen, Carl-Benz-Straße 199, 47057 Duisburg, Germany,}
\author{Ralf Sch\"utzhold}
\affiliation{Fakult\"at f\"ur Physik, Universit\"at Duisburg-Essen, Lotharstra{\ss}e 1, 47057 Duisburg, Germany,}
\affiliation{Helmholtz-Zentrum Dresden-Rossendorf, Bautzner Landstra{\ss}e 400, 01328 Dresden, Germany,}
\affiliation{Institut f\"ur Theoretische Physik, Technische Universit\"at Dresden, 01062 Dresden, Germany.}

\date{\today}

\begin{abstract}
We study relaxation dynamics in a strongly-interacting two-site Fermi-Hubbard model that is induced by coupling each site to a local fermionic bath.
To derive the proper form of the Lindblad operators that enter an effective description of the system-bath coupling in different temperature regimes, we employ a diagrammatic real-time technique for the time evolution of the reduced density matrix.
In spite of a local coupling to the baths, the found Lindblad operators are non-local in space. 
We compare with the local approximation, where those non-local effects are neglected.
Furthermore, we propose an improvement on the commonly-used secular approximation (rotating-wave approximation), referred to as coherent approximation, which turns out superior in all studied parameter regimes (and equivalent otherwise).
We look at the relaxation dynamics for several important observables and compare the methods for early and late times in various temperature regimes. 
\end{abstract}

\maketitle

\section{Introduction}
The description of non-equilibrium dynamics of strongly
interacting many-body systems is a challenging task.
This is already true for closed quantum systems that undergo a unitary time evolution \cite{CC06,MWNM07,IC09,SF11,MK08,EKW09,KLA07,R09,CFMSE08,QKNS14,NS10,KWW06,TCFetal12,GKLetal12,D91,S94,BBW93,RDYO07,RDO08,EHKetal09,CR10,PSSV11,GME11,BCH11,BDK15,KISD11}.
It is even more true for the more realistic scenario of a quantum system coupled to an environment \cite{BP02,RVM13,RWNP19,RW19,SHM08}.
The type of coupling, e.g., whether it allows for exchange of energy and/or spin and/or particles, 
does not only modify the internal quantum dynamics of the system, in many 
cases it will even dominate the system's behavior.

Dynamics of open quantum systems coupled to Markovian baths are generically described by master equations in Lindblad form 
\cite{GKS76,L76,AJ14,HCG13,KGIetal12}.
A common approach is to incorporate the effects of the coupling to the environment in the so-called Lindblad operators on phenomenological grounds 
\cite{QS19}. 
They are introduced ad-hoc, based on physical intuition.
This may work well in some situations but may encounter ambiguities in other cases, in particular for interacting many-body systems.

To obtain a microscopic description of the coupling to the environment one can introduce external baths and their coupling to the system on the Hamiltonian level.
Integrating out the baths yields an effective description of the system's degrees of freedoms in terms of a reduced density matrix.
As an advantage, the microscopic approach does not involve any guessing.
The disadvantage, on the other hand, lies in the increasing complexity of the calculations with increasing size of the many-body quantum system.

The described dichotomy of phenomenological and microscopic approaches asks for trying to combine the best of the two worlds.
As a first step along such an endeavor, we analyze, in the present paper, the Fermi-Hubbard model where each site is coupled separately to its own thermal fermionic bath. For a small model system with only a few interacting degrees of freedom, a full microscopic treatment is feasible. 
To be specific, we choose as a system a Hubbard dimer \cite{CFSB15} and assume that each of the two sites is weakly coupled to a thermal fermionic bath, i.e., electrons can enter from or leave to the bath. 

The first goal of the paper is to intertwine two different theoretical approaches to study non-equilibrium dynamics in the Fermi-Hubbard model coupled to thermal baths: On the one hand, in Sec. \ref{sec:lindblad}, we use a phenomenological method where for the limiting cases of cold $T\ra 0$ and hot $T\ra\infty$ baths the Lindblad operators are deduced from physical intuition. 
On the other hand, in Sec. \ref{sec:diagrammatics}, we formulate the kinetic equations of the system within a real-time technique in where relaxation processes are represented in terms of diagrams. 
Then, in Sec. \ref{sec:connection}, we derive for a single site the Lindblad operators from the respective diagrams in the Markovian limit for the leading order perturbation theory in the system-bath coupling. To overcome the problem of non-positivity, we introduce an alternative (referred to as coherent approximation) to the commonly-used secular approximation. Then, we extend the procedure to systems with more than one site, and we illustrate that the derived Lindblad operators become effectively non-local in space. Only by neglecting those non-local effects we find (for the coherent approximation) an agreement with the phenomenological model. Thus, if non-locality becomes important, the phenomenological approach fails. 
To illustrate our findings, we study, in Sec. \ref{sec:dimer}, a model system that can be treated fully analytically, namely a Hubbard dimer coupled to fermionic baths. 
We compare the various approximations for the limiting cases of a hot and a cold bath by studying the relaxation dynamics for several important observables. 
At last, for small but finite temperatures, we find an additional timescale responsible for the ultra-slow decay of singlet-triplet like spin oscillations. The emergence of different timescales has also been observed in various other systems due to relaxation or thermalization processes \cite{BPGDA09,GJM10,F14,Z15,SJM16,EF16,RHCL13,SPK15}. 

\section{Lindblad master equation}\label{sec:lindblad}

The Fermi--Hubbard model is described by the Hamiltonian \cite{H63}
\begin{equation} \label{eq:Hamiltonian}
  \h H_\text{s} = - J \sum_{\<m,m^\prime\>, \sigma} \h c\s_{m,\sigma} \h c_{m^\prime,\sigma}  + U \sum_m \h n_{m,\sd} \h n_{m,\su} 
  + \eps \sum_{m,\sigma}  \h n_{m,\sigma} ,
\end{equation}
with $\h c\s_{m,\sigma}, \h c_{m,\sigma}$ and $n_{m,\si}$ being the creation, annihilation and occupation-number operators for an electron with spin $\sigma$ at site $m$, respectively. 
The first term describes tunneling between nearest neighbors $\<m,m^\prime\>$ with amplitude $J$, the second term the on-site Coulomb interaction of strength $U$, and $\eps$ is the single-electron energy.
To study relaxation dynamics, we couple the system to external baths and describe the system by the density matrix $\rho(t)$. 
In a generic form, the time evolution is expressed as $\rho(t)=\Pi(t,t_0)\rho(t_0)$ with $\Pi(t,t_0)$ being a dynamical map propagating the reduced density matrix forward in time. By imposing Markovian dynamics in form of a \textit{semigroup} assumption for the propagator, $\Pi(t,t^\prime)\Pi(t^\prime,t_0)=\Pi(t,t_0)$ with $t_0<t^\prime<t$, it can be shown that the dynamics of $\rho(t)$ are described by a Lindblad equation \cite{GKS76,L76} $\dot{\rho} = \mathcal{L} \h\rho$ of the form 
\begin{align} \label{eq:Lindbladform1}
  \dot{\rho} = -\frac{i}{\hbar}\, [\h \tilde{H}_\text{s},\rho]+\sum_{\mu,\nu=1}^{N^2{-}1}\ga_{\mu\nu}
  \left(\h \Eb_{\mu}\, \h\rho\, \h \Eb\s_{\nu} - \frac1{2}\{\h \Eb\s_{\nu}\,\h \Eb_{\mu}, \h\rho\}\right),
\end{align}
where the solution can formally be given as $\Pi(t,t_0)=e^{\mathcal L (t-t_0)}$. The generator $\L$ of the semigroup is called \textit{Liouville superoperator} or simply \textit{Liouvillian}. In Eq. \eqref{eq:Lindbladform1}, $\Eb_\mu$ with $\mu=1,..,N^2{-}1$ span a basis for the operators defined in the system's Hilbert space of dimension $N$. We choose $\Eb_{N^2}\propto\1$ (which has no effect for the time evolution), leaving $N^2-1$ independent operators $\Eb_\mu$.
The $N^2-1$-dimensional matrix $\ga_{\mu\nu}$ describing the coupling to the baths is hermitian and positive semidefinite. Then, per construction, the Lindblad equation preserves the trace $\text{tr}\rho(t)=1$, hermiticity $\rho(t)\s=\rho(t)$ and positivity $\rho(t)\geq 0$, all properties which are essential for a physical interpretation of the density matrix $\rho(t)$. Note that, in general, we have $\tilde{H}_\text{s} \neq H_\text{s}$, i.e. a modified coherent evolution due to renormalization effects induced by the baths. Using that $\ga_{\mu\nu}=\ga_{\nu\mu}^*$ is hermitian, we 
diagonalize the Lindblad equation arriving at 
\begin{align} \label{eq:Lindbladform2}
  \dot{\rho} = -i\, [\h \tilde{H}_\text{s},\rho]+\sum_{k=1}^{N^2-1}\ga_{k}
  \left(\h \ell_{k}\, \h\rho\, \h \ell\s_{k} - \frac1{2}\{\h \ell\s_{k}\,\h \ell_{k}, \h\rho\} \right) ,
\end{align}
with (from now on) natural units $\hbar=1$. In Eq. \eqref{eq:Lindbladform2}, $\ell_{k}$ with $k=1,...,N^2{-}1$ are referred to as \textit{Lindblad} operators describing the interaction with the environment (e.g. gain and loss terms). This diagonal Lindblad form will be the starting point of the phenomenological description of the Hubbard system coupled to thermal baths. Here, without knowing the microscopic details of the environment, we incorporate the Lindblad operators $\ell_{k}$ and the coupling parameters $\ga_k\geq 0$ solely from the system's perspective in such a way that they model the desired effects.

\begin{figure}[ht]
  \includegraphics[width=0.75\linewidth]{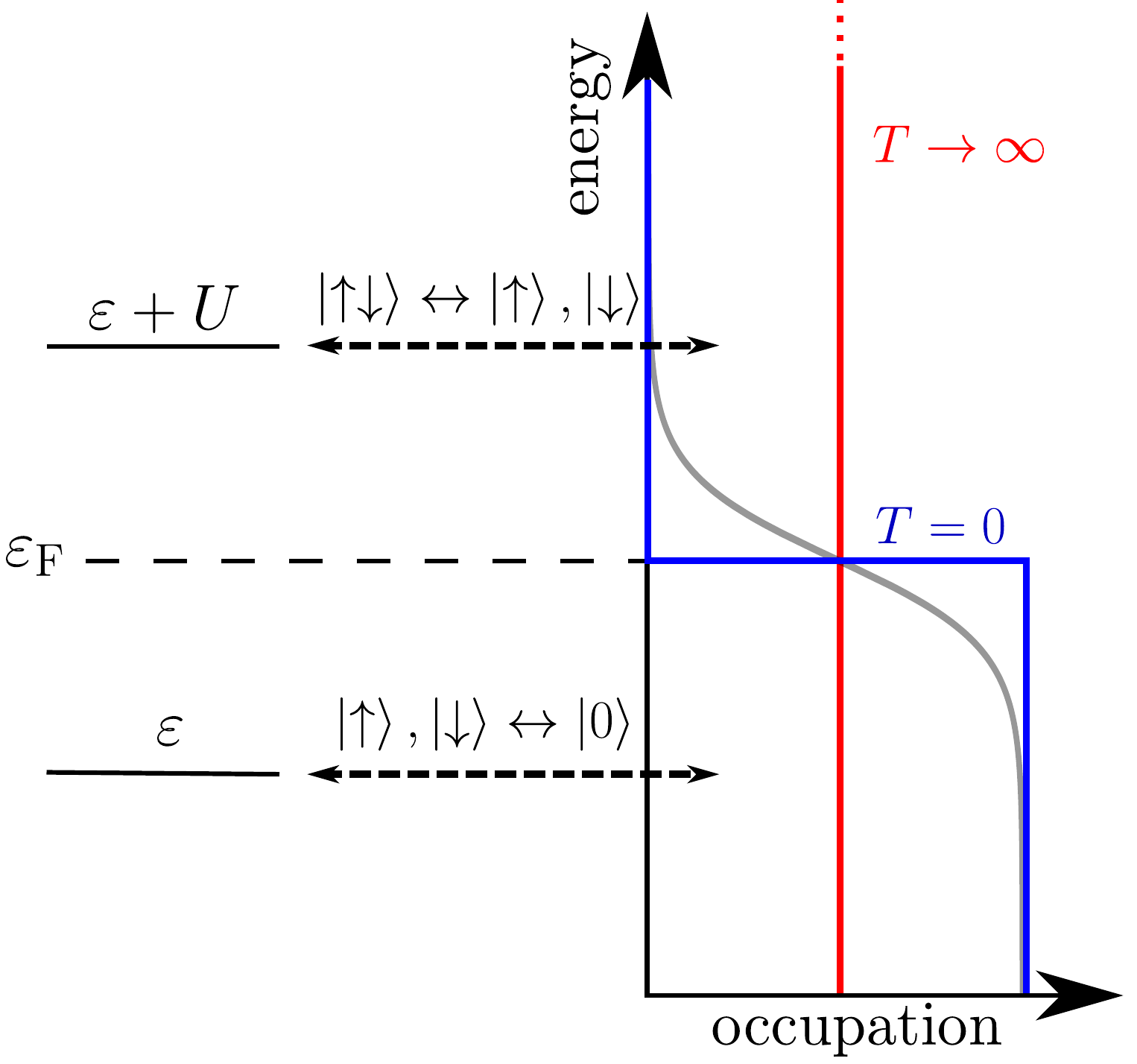}
  \caption{Single site coupled to a hot (red), cold (blue) and finite-temperature (gray) bath. Electron-electron interaction $U$ gives rise to two excitation energies $\eps$ and $\eps+U$.}
  \label{fig:bath}
\end{figure}

In this paper, we study a situation where every site $m$ is coupled individually to its own bath (one site is shown in Fig. \ref{fig:bath}). The baths are assumed to be identical, i.e., they have the same temperature $T$ and the same Fermi energy $\eps_\text{F}$. In the phenomenological model, we assume that given a local coupling, the Linblad operators  for each site $m$ depend only on locally defined operators.  Specifically, the Hilbert-space dimension of one site is $N=4$ and, therefore, the local Lindblad operators are constructed from 16 linear independent operators.
Half of them, namely $\{\1,\h c_{\su}\h c_{\sd},\h c\s_{\su}\h c\s_{\sd},\h c\s_{\si}\h c_{\bar\si},\h n_{\si},\h n_{\su} \h n_{\sd}\}_m$, where $\bar\si$ denotes the opposite spin to $\si$, contain an even number of Fermi operators.
The other half, $\{\h c_\si, \h c\s_\si,\h c_\si\h n_{\bar\si}, \h c\s_\si\h n_{\bar\si}\}_m$, contains an odd number of Fermi operators.
The Lindblad operators constructed from the second class change the Fermion parity number $P_m=(-1)^{n_m}$ with the local occupation number $n_m=\sum_\sigma n_{m,\si}$ on site $m$.
Therefore, we refer to a bath described by these Lindblad operators as a \textit{fermionic} bath.
In contrast, the Lindblad operators of the first class leave the Fermion parity number unchanged, describing a \textit{bosonic} bath.
(Note that the terms bosonic and fermionic do not necessarily refer to the actual nature of the particles involved in the physical bath.)
In general, both types of baths can exchange charge, spin and energy with the system and thus relaxation and decoherence dynamics can be studied. 
In Ref. \onlinecite{QS19}, bosonic baths with the Lindblad operators $\h n_{m,\si}$ have been considered.
Here, we investigate fermionic baths only. Considering separate channels where single electrons of spin $\sigma$ ($\uparrow$ or $\downarrow$) enter ($\alpha=+$) or leave ($\alpha=-$) the site $m$, we are left with at maximum two independent Lindblad operators $(\ell^\alpha_{m,\sigma})_k$ with $k=1,2$ constructed from operators of the fermionic type $c_{m,\si}^\alpha$ and $\h c_{m,\si}^\alpha \h n_{m,\bar\si}$, where $c_{m,\si}^+\equiv c\s_{m,\si}$ and $c_{m,\si}^-\equiv c_{m,\si}$. 
Moreover, we assume $(\ga^\alpha_{m,\sigma})_k=(\ga^\alpha)_k$ meaning that the coupling is identical at every site and symmetric with respect to spin.  

In simple cases, the explicit form of the Lindblad operators can be written down using phenomenological arguments. In the following we will discuss two such examples of cold ($T\rightarrow 0$) and hot ($T\rightarrow\infty$) baths before comparing it to a in--depth microscopic analysis. 

\subsection{Hot baths}

We assume each site $m$ is coupled independently to a (separate) hot fermionic bath (cf. red line in Fig. \ref{fig:bath}). Only single-electron excitations of the form $\ket{\si} \leftrightarrow \ket{0}$ and $\ket{\uparrow\downarrow}\leftrightarrow  \ket{\si}$ with excitation energies $\eps$ and $\eps+U$ are considered. In the limit $T\rightarrow\infty$, each bath is assumed to have free and occupied states available at every energy so that the details of the system's energy spectrum as well as the occupation become dispensable. 
This motivates to choose the Lindblad operators $\ell^\alpha_{m,\si}$ for adding or removing an electron with spin $\si$ to be independent of the number $\h n_{m,\bar\si}$ of electrons with opposite spin.
This yields
\begin{align}
  \h \ell^+_{m,\si} &= \, \h c\s_{m,\si}, & \h \ell^-_{m,\si} &=\, \h c_{m,\si},  \label{eq:lhot}
\end{align}
with couplings $\ga_\text{hot}^+$ and $\ga_\text{hot}^-$ for injecting or removing a particle, respectively. 
Notice that the choice of only one of two possible Lindblad operators for each $m,\si$ and $\alpha$ is sufficient and therefore we have dropped the index $k$ here. 

\subsection{Cold baths}

In the opposite limit, each site $m$ is coupled to a cold fermionic bath, formally at zero temperature. 
The states in each bath are fully occupied up to the Fermi energy $\eps_\text{F}$ and hence Pauli--blocked. 
Above $\eps_\text{F}$ the states are empty (cf. blue line in Fig. \ref{fig:bath}). 
The choice $\eps < \eps_\text{F} < \eps + U$ introduces a natural filter for the effective couplings between the site and the bath: 
for empty sites $\ket{0}$ electrons can only enter (transferred energy $\eps < \eps_\text{F}$),
for doubly occupied sites $\ket{\uparrow\downarrow}$ the electrons can only leave (transferred energy $\eps + U > \eps_\text{F}$)
while for singly-occupied sites no electron transfer is possible as both channels are blocked.  
Therefore, we assume the following occupation-dependent Lindblad operators
\begin{align}
  \h \ell^+_{m,\si} &= \, \h c\s_{m,\si}\, (1-\h n_{m,\bar\si}), &  \h \ell_{m,\si}^-  &=\,  \h c_{m,\si}\, \h n_{m,\bar \si}, \label{eq:lcold}
\end{align}
which inject an electron with rate $\ga_\text{cold}^+$ if the site is empty and remove it with rate $\ga_\text{cold}^-$ if it is doubly occupied, respectively. 
Again, as we deal with only one Lindblad operator per $m,\si$ and $\alpha$, we drop the index $k$.

\subsection{Finite-temperature baths}

At finite temperature $T$ (cf. gray line in Fig. \ref{fig:bath}), the relaxation channels are neither treated all equally (as for $T\rightarrow \infty$) nor are they fully blocked in one direction (as for $T\rightarrow 0$). One possible ansatz for a finite-temperature bath is to interpolate between the hot and cold cases to get a single Lindblad operator 
\begin{align}
  \h \ell^+_{m,\si}&=g_+(T) c\s_{m,\si}\,\h (1-n_{m,\bar\si}) + h_+(T)\h c\s_{m,\si}\, , \nonumber\\
  \h \ell^-_{m,\si}&=g_-(T) c_{m,\si}\,\h n_{m,\bar\si} + h_-(T)\h c_{m,\si}\, , \label{eq:lfinite}
\end{align} 
with temperature- (and energy-) dependent functions $g_\alpha(T)$ and $h_\alpha(T)$.  Demanding that $g_\alpha(T\rightarrow\infty)=0$ and $h_\alpha(T\rightarrow0)=0$, we ensure that the Lindblad operators give Eq. \eqref{eq:lhot} and Eq. \eqref{eq:lcold} for $T\rightarrow \infty$ and $T\rightarrow 0$, respectively. To obtain the resulting Liouvillian ${\cal L}_\text{phen}$ of the phenomenological approach, we have to insert the Lindblad operators into Eq. \eqref{eq:Lindbladform2}. 
Note, that in the case of finite temperatures, it is less intuitive to anticipate the proper form of the Lindblad operators as well as to decide whether one or two Lindblad operators per $m$, $\sigma$ and $\alpha$ should be used. 
This issue will be discussed in the next section, where we start from a microscopic description of the baths. 

\section{Real-time diagrammatics} \label{sec:diagrammatics}

In this part, we give a brief introduction to the real-time diagrammatic approach \cite{KSSS96,KSS96}, with the goal to give the so-far phenomenological discussion a microscopic foundation. 
We will proceed in the following way: We start from a fully-microscopic description of system and environment, and then employ the real-time diagrammatic technique to arrive at a formally exact kinetic master equation for the reduced density matrix of the system by tracing out the environment. The diagrammatic representation is, then, an ideal starting point for a systematic perturbation expansion in the system-bath coupling.

The total Hamiltonian
\begin{align}
H_\text{tot}=H_\text{s}+H_\text{b}+H_{\text{c}}
\label{eq:Hdot}
\end{align}
consists of three parts, with s, b and c denoting the system, the bath and the coupling, respectively.
The system $H_\text{s}$ is the Hubbard Hamiltonian of Eq. \eqref{eq:Hamiltonian} with eigenvectors and eigenenergies denoted by $\ket{\chi}$ and $E_\chi$, respectively. The environment $H_\text{b}$, on the other hand, consists of independent, but identical fermionic baths at every site $m$
\begin{align}
H_{\text{b}}&=\sum_{k,\sigma} \eps_{k,\sigma} a^{\dagger}_{m,k,\sigma} a_{m,k,\sigma}\, ,
\end{align} 
where $a^\dagger_{m,k,\sigma}$ and $a_{m,k,\sigma}$ are creation and annihilation operators for lead electrons at site $m$ with orbital and spin degrees of freedom $k$ and $\sigma$. We assume the electrons are noninteracting and $\eps_{k,\sigma}$ is the single-particle energy. At last, the tunnel coupling between the system and the environment is described by
\begin{align}
H_{\text{c}}=\sum_{m,k,\sigma } t_{k,\sigma} c_{m,\sigma}^{\dagger} a_{m,k,\sigma} +\text{h.c.}\, , 
\end{align}
where $t_{k,\sigma}$ is the (site-independent) tunneling amplitude for an electron entering or leaving the site $m$. Note that in $H_{\text{c}}$ different sites $m$ and $m^\prime$ do not mix, which ensures a local coupling. This coupling Hamiltonian $H_{\text{c}}$ will later be treated as a perturbation and it will turn out that only even powers $H_{\text{c}}^{2n}$, with $n$ being a nonnegative integer, will contribute. 

To determine the time evolution of the reduced density matrix $\rho_{\chi_2}^{\chi_1}(t)$, we have to calculate the expectation values of the dyads $\dyad{\chi_2}{\chi_1}$. 
We change to the interaction picture with respect to $H_{\text{c}}$,
\begin{align}
\rho_{\chi_2}^{\chi_1}(t)=\text{tr} \left\lbrace T_{\mathcal K} \left[ e^{-i \int_{\mathcal K} \mathrm{d}t^\prime H_{\text{c},\text{I}}(t^\prime)}\dyad{\chi_2}{\chi_1}_\text{I}(t) \right] \rho^\text{tot}_0 \right\rbrace,
\end{align} 
and, then, formally expand in the coupling,
\begin{align}
\rho_{\chi_2}^{\chi_1}(t)&=\sum_{n=0}^\infty \underbrace{\int_{\mathcal K} \mathrm{d}t_1 \int_{\mathcal K} \mathrm{d}t_2 ...\int_{\mathcal K} \mathrm{d}t_n}_{t_1<t_2<...<t_n}  \left(-i\right)^n \times \nonumber \\ &\text{tr} \bigg \lbrace T_{\mathcal K} \left[   H_{\text{c},\text{I}}(t_1) H_{\text{c},\text{I}}(t_2)... H_{\text{c},\text{I}}(t_n) \dyad{\chi_2}{\chi_1}_\text{I}(t) \right]  \rho^\text{tot}_0 \bigg \rbrace \,, \label{eq:observableA}
\end{align} 
where the integrals are evaluated along the Keldysh contour ${\mathcal K}$, i.e. it runs first forwards in time from $t_0$ to $t$ and then backwards from $t$ to $t_0$. The Keldysh time-ordering operator $T_{\mathcal K}$ orders operators such that earlier times (with respect to the Keldysh contour) appear further to the right. The index I indicates the interaction picture $A_\text{I}(t)=e^{i(H_{\text{s}}+H_{\text{b}})(t-t_0)}A(t)e^{-i(H_{\text{s}}+H_{\text{b}})(t-t_0)}$. Moreover, we assume that at time $t_0$, the system and the bath are in a product state $ \rho^\text{tot}_0 =\rho_0 \otimes \rho_0^\text{b}$, whereas the bath is assumed to be in a thermal Gibbs state $\rho_0^\text{b}=e^{-\beta (H_\text{b}-\eps_\text{F} N)}/Z$ with inverse temperature $\beta=(k_\text{B}T)^{-1}$, chemical potential $\eps_\text{F}$ and partition sum $Z=\tr e^{-\beta (H_\text{b}-\eps_\text{F} N)}$. Note that the factor $1/n!$ from the  Taylor expansion cancels with the $n!$ possible permutations of $t_1,t_2, \ldots, t_n$. 
Performing the partial trace over the bath and making use of the fact that $H_\text{b}$ is bilinear in the creation and annihilation operators, we can apply Wick's theorem to contract the lead electron operators of $H_{\text{c},\text{I}}(t_{n^\prime})$ with $n^\prime=1, \ldots, n$ pairwise in all possible ways and replace them by bath equilibrium distributions $\ev{a^{\dagger}_{m,k,\sigma}a_{m,k,\sigma}}=f_+(\eps_{k,\sigma})$ and $\ev{a_{m,k,\sigma}a^{\dagger}_{m,k,\sigma}}=f_-(\eps_{k,\sigma})$, where we use the notation $f_+\equiv f$ and $f_-\equiv 1-f$ with $f(\omega)=[e^{\beta (\omega{-}\eps_\text{F})}{+}1]^{-1}$ being the Fermi-Dirac distribution. After employing Wick's theorem we can visualize the Eq. \eqref{eq:observableA} in the following way
\begin{widetext}
\begin{align}
\rho_{\chi_2}^{\chi_1}(t)=\sum_{\chi_1^\prime,\chi_2^\prime}\sum_{n=0}^\infty  \sum_{\substack{\text{{\footnotesize{all}}} \\ \text{\footnotesize{contractions}}}}  \underbrace{\int_{\mathcal K} \mathrm{d}t_1 \int_{\mathcal K} \mathrm{d}t_2 ...\int_{\mathcal K} \mathrm{d}t_{2n}}_{t_1<t_2<...<t_{2n}}  \raisebox{-2mm}{\includegraphics[scale=0.32, angle=0,valign=M]{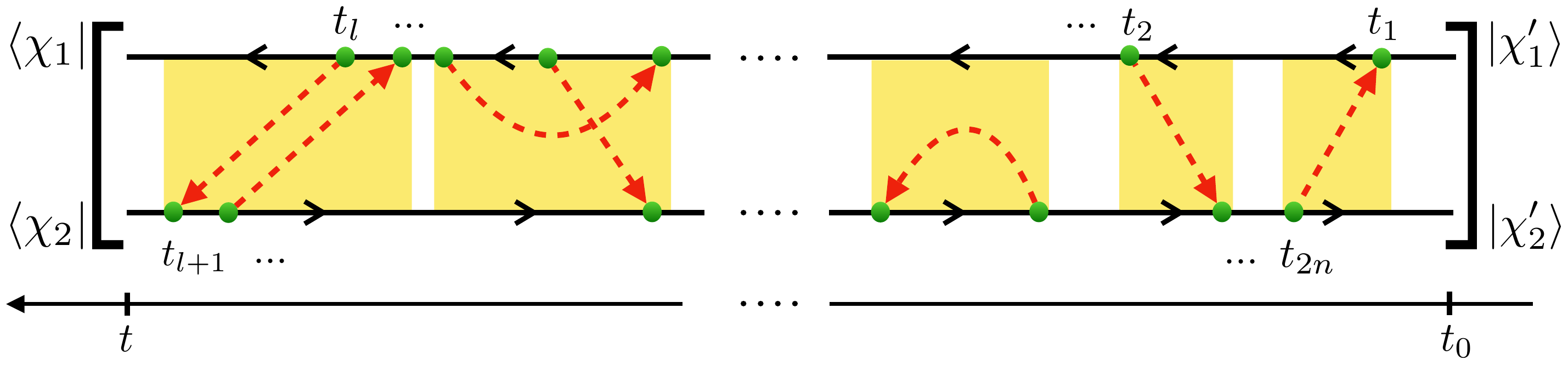}}  \rho^{\chi_1^\prime}_{\chi_2^\prime}(t_0), \label{eq:propagator}
\end{align} 
\end{widetext}
where we have evaluated the trace of the system's degrees of freedom over $\ket{\chi_2^\prime}$ and inserted $\1=\sum_{\chi_1^\prime}\dyad{\chi_1^\prime}{\chi_1^\prime}$ right behind $\rho_0$ to obtain $\rho^{\chi_1^\prime}_{\chi_2^\prime}(t_0)=\mel{\chi_1^\prime}{\rho_0}{\chi_2^\prime}$. In Eq. \eqref{eq:propagator}, the Keldysh time increases anticlockwise from $t_1$ at the upper right corner to $t_{2n}$ at the lower right corner of the contour, whereas the real time simply increases from right $t_0$ to left $t$. The vertices (green dots) represent the operators $c^\dagger_{m,\sigma}$ and $c_{m,\sigma}$ and the tunneling lines (red dashed lines) diagrammatically represent the contractions of the lead electron operators $a^{\dagger}_{m,k,\sigma}$ and $a_{m,k,\sigma}$ and they indicate that an electron is entering (pointing towards the contour) or leaving (pointing away from the contour) the system. The contour line itself indicates the present state of the system. Note, that we have to consider only even 
orders $2n$ since, according to Wick's theorem, contractions with an odd number of lead electron operators vanish. For the same reason, the numbers of creation and annihilation operators have to be equal. \\ \newline
We can formally abbreviate Eq. \eqref{eq:propagator} as 
\begin{align}
&\rho_{\chi_2}^{\chi_1}(t)=\sum_{\chi_1^\prime,\chi_2^\prime} \Pi_{\chi_2 \chi_2^\prime}^{\chi_1 \chi_1^\prime}(t,t_0) \rho^{\chi_1^\prime}_{\chi_2^\prime}(t_0)\nonumber \\
\Leftrightarrow \quad &\rho(t)=\Pi(t,t_0)\rho(t_0), \label{eq:propagator2}
\end{align}
where in the second line we recast the formula into an equation with supervectors $\rho(t)$ and superoperators $\Pi(t,t_0)$ defined in Liouville space by employing the Hilbert-Schmidt scalar product $\left(A,B\right)=\text{tr} [A^\dagger B]$ for linear operators $A$ and $B$. Explicitly, we used the connection $\rho_{\chi_2}^{\chi_1}=\left(\dyad{\chi_1}{\chi_2},\rho\right)$ and $\Pi_{\chi_2 \chi_2^\prime}^{\chi_1 \chi_1^\prime}=\left(\dyad{\chi_1}{\chi_2},\Pi \dyad{\chi_1^\prime}{\chi_2^\prime}\right)$ to arrive at an equation free of indices. The superoperator $\Pi(t,t_0)$ is a propagator and describes the time evolution of the reduced density matrix from $t_0$ to $t$. It fulfills a Dyson equation of the form
\begin{align}
\Pi(t,t^\prime)=&{\Pi^{(0)}}(t,t^\prime) + \nonumber   \\ &\int\limits_{t^\prime}^t \mathrm{d}t_2 \int\limits_{t^\prime}^{t_2}  \mathrm{d}t_1  \Pi(t,t_2)\W(t_2,t_1) {\Pi^{(0)}}(t_1,t^\prime), \label{eq:dyson}
\end{align}
where ${\Pi^{(0)}}(t^\prime,t) =e^{\L_0(t-t^\prime)}$ with $\L_0=-i \lbrack H_\text{s},\cdot\rbrack$ describes the free propagation without any tunneling and $\W(t_2,t_1)$ is the irreducible self-energy, defined as the sum of all topologically inequivalent diagrams that cannot be cut vertically into smaller diagrams.
In Eq. \eqref{eq:propagator}, the irreducible diagrams are highlighted with a yellow background. To calculate the irreducible diagrams in every order $2n$, which give rise to the components of  $\W(t_2,t_1)$, we simply have to follow the diagrammatic rules: 
\begin{enumerate}
\item Draw all irreducible and topological inequivalent diagrams with $n$ directed tunneling lines that connect pairwise all $2n$ vertices.
\item Every vertex where the tunneling line points towards (away from) the contour corresponds to a transition matrix element $\mel{\chi}{c^{\dagger}_{m,\si}}{\chi^\prime}$ ($\mel{\chi}{c_{m,\si}}{\chi^\prime}$) with $\ket{\chi^\prime}$ being the state prior to $\ket{\chi}$ with respect to the Keldysh time.  
\item Every directed tunneling line pointing from $t^\prime$ to $t$ induces a factor $\Gamma_\si(\omega) f_\alpha(\omega)e^{- i \omega (t-t^\prime)}/(2\pi)$ with $\alpha=\pm$ for tunneling lines aligned with ($-$) or against ($+$) the Keldysh contour.
\item Each free propagating segment on the Keldysh contour with state $\ket{\chi}$ between $t^\prime$ and $t>t^\prime$ implies a factor $e^{-i E_\chi (t-t^\prime)}$ on the upper and $e^{i E_\chi (t-t^\prime)}$ on the lower contour.
\item Add a factor $(-1)^{a+b+n}$ with $a$ being the number of crossings of tunneling lines (manifestation of the Pauli exclusion principle) and $b$ being the number of vertices on the lower contour.
\item Integrate over intermediate times (respecting the ordering) and sum over lead degrees of freedom (energy $\omega$ and spin $\si$).
\end{enumerate}
To formulate the rules, it was beneficial  --- since the bath is macroscopic --- to switch from the orbital degree of freedom $k$ to energy $\omega$ by replacing formally $2\pi\sum_k \vert t_{k,\si}\vert^2 \rightarrow \int_{-\infty}^\infty \mathrm{d}\omega\, \Ga_\si(\omega)$ and thereby introducing the tunnel-coupling strength as $\Gamma_\sigma(\omega):=2\pi \sum_{k} \vert t_{k,\sigma} \vert^2 \delta(\omega-\eps_{k,\sigma})$ which conveniently serves as a parameter denoting the order $n$ in perturbation theory ${\cal O}(\Ga^n)$. \\
Now, we have all ingredients to write down a generalized master equation for $\rho(t)$. 
By differentiating Eq. \eqref{eq:propagator2} and inserting the propagator of Eq. \eqref{eq:dyson} we arrive finally at a formally exact kinetic equation for the reduced density matrix
\begin{align}\label{eq:kineticeq}
\dot{\rho}(t)=-i \left[ H_\text{s},\rho(t)\right] + \int\limits_{t_0}^t \mathrm{d}t^\prime{\mathcal W}(t{-}t^\prime)\rho(t^\prime)\, ,
\end{align}
where the irreducible self-energy takes the role of a transition matrix in Liouville space. Note that due to translational invariance in time we have ${\mathcal W}(t,t^\prime)={\mathcal W}(t-t^\prime)$. 
So far, the equation is formally still exact if all (infinitely many) irreducible diagrams are included. However, to arrive at a Markovian Lindblad equation, several approximations are necessary \cite{BP02}, which will be discussed in the next section.

\section{From Diagrammatics to Lindblad}\label{sec:connection}

To cast the exact kinetic equation \eqref{eq:kineticeq} into the Lindblad form, we will employ several approximations on $\W(t{-}t^\prime)$. Most importantly, we will only be interested in a weak coupling between system and bath. Therefore, perturbation theory to leading order ${\cal O}(\Ga)$ is sufficient for our description and we write ${\mathcal W}(t-t^\prime)\approx {\mathcal W}^\text{(1)}(t-t^\prime)$.
The index $(1)$ denotes that only irreducible diagrams with one tunneling line are considered (three examples are indicated with a yellow background in Eq. \eqref{eq:propagator}). In zeroth order the density matrix in the interaction picture is time-independent $\rho_\text{I}(t^\prime)\approx \rho_{\text{I}}(t)$, and thus we replace $\rho(t^\prime)\approx e^{ -{\mathcal L}_0 (t-t^\prime)}\rho(t)$ in the integral of Eq. \eqref{eq:kineticeq} to obtain a time-local master equation 
\begin{align}
\dot{\rho}(t)=-i \left[ H_\text{s},\rho(t)\right] +\left[ \int\limits_{0}^{t-t_0} \mathrm{d}\tau{\mathcal W}^\text{(1)}(\tau)e^{ -{\mathcal L}_0 \tau}\right]\rho(t)\, ,\label{eq:timelocal}
\end{align}
that is consistent with a leading-order perturbation expansion. In Eq. \eqref{eq:timelocal} we made use of the substitution $\tau=t{-}t^\prime$.
Note that the assumption of a weak coupling to the environment is only justified, if the irreducible diagrams of ${\mathcal W}^\text{(1)}(\tau)$ have a negligible width (duration of a tunneling event) compared to their mean distance (time between tunneling events). Only then, first-order diagrams do not overlap in time and hence there is no necessity of higher-order corrections.
Employing the diagrammatic rules, every first-order diagram in ${\mathcal W}^\text{(1)}(\tau)$ gives rise to a transition rate of the form 
\begin{align}
\ga_{\pm,\si}^\text{(1)}(\tau)&=\Re\left[\int\limits_{-\infty}^\infty \frac{\mathrm{d}\omega}{2\pi}\Ga_\si(\omega) f_\pm(\omega)e^{i (\omega{-}\Delta E)\tau} \right]  \nonumber \\
&=\frac{\Ga}{4} \left\{\delta(\tau{-}0^+) \mp 2 k_\text{B} T \frac{\sin \left[( \Delta E-\eps_\text{F}) \tau\right]}{\sinh(\pi k_\text{B} T \tau)} \right\},
\label{eq:transitionrate}
\end{align}
with a specific excitation energy $\Delta E$ of the system. To arrive at the second line of Eq. \eqref{eq:transitionrate}, we assumed the wide-band limit $\Gamma_\si(\omega)=\Gamma_\si$ as well as spin-independent tunneling rates $\Ga_\si=\Ga$. We find that the characteristic decay time $\tau_\text{c}$ of $\ga_{\pm,\si}^\text{(1)}(\tau)$ (or the width of a diagram) is given by the smaller value of $\hbar/\vert \Delta E{-}\eps_\text{F}  \vert_\text{min}$ and $\hbar/(k_\text{B}T)$ (see also Ref. \onlinecite{T08}). The mean distance between diagrams, on the other hand, is given by $\hbar/\Gamma$. Hence, if either $\Gamma \ll  \vert \Delta E{-}\eps_\text{F} \vert$ (all excitation energies $\Delta E$ are far away from the Fermi energy) or $\Gamma \ll k_\text{B}T$ (temperature is sufficiently high) is fulfilled, the approximation of a weak coupling is justified.
Under those conditions, we can send $t_0\rightarrow -\infty$ in the integral of Eq. \eqref{eq:timelocal}, and by making use of $\int_{0}^\infty \mathrm{d}\tau \ga_{\pm,\si}^\text{(1)}(\tau) = \Ga\frac{f_\pm(\Delta E)}{2}$, we effectively arrive at the Markovian limit. We define the time-independent transition matrix as ${\mathcal W}^\text{M}=\int_0^{\infty}\mathrm{d}\tau {\mathcal W}^\text{(1)}(\tau)e^{-{\mathcal L}_0\tau}$, where M denotes the Markov assumption. (At this point the master equation is identical to the Bloch-Redfield equation \cite{WB53,R57,T08}.) 
Note that, in this paper we consider only the real part of the diagrams (see Eq. \eqref{eq:transitionrate}), so that $\W^\text{M}$ becomes purely real and renormalization effects (see Ref. \onlinecite{SGK12}) are neglected. This corresponds to setting $\tilde{H}_\text{s}=H_\text{s}$ in Eq. \eqref{eq:Lindbladform2}.

At this stage, the Markovian master equation is generally not yet in the Lindblad form and, thus, does not guarantee positivity for all density matrices $\rho(t)$ evolving according to it. To see this explicitly, as a next step, we consider in detail the case of a single-site Hubbard Hamiltonian coupled to an electronic reservoir. Afterwards, we discuss the generalization to the case of $M$ sites.

\subsection{One site}\label{subsec:onesite}

Trying to bring ${\mathcal W}^\text{M}$ into Lindblad form, it is beneficial to group its components as 
\begin{align}
\left({{\mathcal W}^\text{M}}\right)_{\chi_2\chi_2^\prime}^{\chi_1 \chi_1^\prime}=\sum_{\alpha,\sigma}\left({\mathcal W}_{\sigma}^{\text{M},\alpha}\right)_{\chi_2\chi_2^\prime}^{\chi_1 \chi_1^\prime}, 
\end{align}
i.e., we split the processes according to the spin $\sigma=\uparrow,\downarrow$ and the direction of the tunneling line (either directed with ($\alpha=-$) or against ($\alpha=+)$ rising Keldysh time). For those directed against the Keldysh contour, we find in total four irreducible diagrams
\begin{widetext} 
  \begin{align}
    \left({\mathcal W}_{\sigma}^{\text{M},+}\right)_{\chi_2\chi_2^\prime}^{\chi_1 \chi_1^\prime}&=\int\limits_0^{\infty}\mathrm{d}\tau\, e^{i(E_{\chi_1^\prime}-E_{\chi_2^\prime})\tau} \raisebox{1mm}{\includegraphics[scale=0.40, angle=0,valign=M]{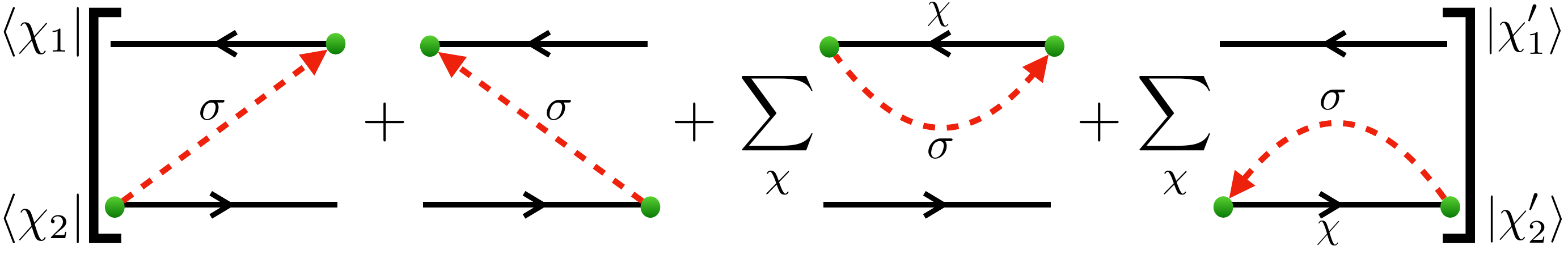}}\nonumber \\
    &=\left(\dyad{\chi_1}{\chi_2},\sum_{\mu,\nu=1}^2 \ga^+_{\mu\nu}\left[\Eb^{+}_{\sigma,\mu} \dyad{\chi_1^\prime}{\chi_2^\prime} \Eb^{+\dagger}_{\sigma,\nu}-\frac{1}{2} \left\{ \Eb^{+\dagger}_{\sigma,\nu}\Eb^+_{\sigma,\mu},\dyad{\chi_1^\prime}{\chi_2^\prime}\right\} \right]\right).\label{eq:diagramstolindblads}
  \end{align}
\end{widetext}
Here, we applied the diagrammatic rules to bring the equation into a form (second line) that formally looks like the matrix elements of the dissipative part of the Lindblad equation \eqref{eq:Lindbladform1}. The used basis operators $\Eb^{+}_{\sigma,1}=c\s_{\sigma}(1{-}n_{\bar \sigma})$ and $\Eb^{+}_{\sigma,2}=c\s_{\sigma} n_{\bar \sigma}$ describe the excitation of the first (site is initially empty) and the second (site is initially singly occupied) electron, respectively. After a lengthy but straightforward calculation, one can identify the first two diagrams with Lindblad terms of the form $\Eb^+_{\si,\mu} \rho {\Eb^{+\dagger}_{\si,\nu}}$, the third one with $\rho {\Eb^{+\dagger}_{\si,\nu}}\Eb^+_{\si,\mu}$ and the last one with ${\Eb^{+\dagger}_{\si,\nu}} \Eb^+_{\si,\mu} \rho$. The coefficients $\ga^+_{\mu\nu}$ can then be given in the matrix form
\begin{align}
  \left(\ga^+_{\mu\nu}\right)=\Ga \begin{pmatrix}
  f_+(\eps) & \frac{f_+(\eps){+}f_+(\eps{+}U)}{2} \\
  \frac{f_+(\eps){+}f_+(\eps{+}U)}{2} & f_+(\eps{+}U)
  \end{pmatrix}.\label{eq:couplingmatrix}
\end{align}
Analogously, we can write down the matrix elements for tunneling lines that are directed with ($\alpha=-$) Keldysh time. 
Then, we arrive at a similar expression with $\Eb^-_{\sigma,\mu}=\Eb^{+\dagger}_{\sigma,\mu}$ and coefficients $\ga^-_{\mu\nu}$ which differ from $\ga^+_{\mu\nu}$ only by the exchange of $f_+\leftrightarrow f_-$.
The master equation reads now
\begin{align}
  &\dot{\rho}(t) = -i\left[ H_\text{s},\rho(t)\right] + \nonumber \\ 
  &\sum_{\alpha,\sigma,\mu,\nu}\ga^\alpha_{\mu\nu}\left[\Eb^\alpha_{\sigma,\mu} \rho(t) \Eb^{\alpha\dagger}_{\sigma,\nu}-\frac{1}{2} \left\{ \Eb^{\alpha\dagger}_{\sigma,\nu}\Eb^\alpha_{\sigma,\mu},\rho(t)\right\} \right],
\end{align}
which has a similar form as equation \eqref{eq:Lindbladform1}. Here, however, $\ga^\alpha_{\mu\nu}$ is not a positive semi-definite matrix and, thus, the positivity (to be more precise non-negativity) of the density matrix is not guaranteed per construction. This can easily be checked by calculating the eigenvalues of $\ga^\alpha_{\mu\nu}$,
\begin{align}
  \ga^\alpha_\pm=\Ga\left(\frac{f_\alpha(\eps)+f_\alpha(\eps{+}U)}{2}\pm \sqrt{\frac{f^2_\alpha(\eps)+f^2_\alpha(\eps{+}U)}{2}}\right),\label{eq:eigenvalues}
\end{align}
where $\ga^\alpha_-\leq 0$. This is a well known problem \cite{ABF07,DTAVB18,FG19,TB15,W08,EKCLK16,SZB09,SKT16}. Here, we discuss two different approaches to deal with it.

\subsubsection{Secular approximation}

\def\LS{{\cal L}^\text{S}}

The usual solution to guarantee positivity is provided by the so-called secular approximation (also termed rotating-wave approximation) \cite{D74,D76,DS79}
\begin{align}
{\mathcal W}^\text{S}=\lim_{T\rightarrow \infty} \frac{1}{2T} \int\limits_{-T}^{T}\mathrm{d}t e^{-{\mathcal L}_0 t} {\mathcal W}^\text{M}e^{ {\mathcal L}_0 t},
\end{align}
where $e^{-{\mathcal L}_0 t} {\mathcal W}^\text{M}e^{ {\mathcal L}_0 t}$ transforms the superoperator $\W^\text{M}$ into the interaction picture. That approximation gets rid of all terms oscillating in the co-rotating reference frame (non-secular terms) by performing a long-time average. 
In order to understand the consequences of the approximation for the diagrams, we look at specific matrix elements 
\begin{align}
&\left( {\mathcal W}^\text{S}\right)_{\chi_2^\prime \chi_2}^{\chi_1^\prime \chi_1}=\text{tr}\left[ (\dyad{\chi_1^\prime}{\chi_2^\prime})^\dagger {\mathcal W}^\text{S} \dyad{\chi_1}{\chi_2} \right]\nonumber\\
&=  \lim_{T\rightarrow \infty} \frac{1}{2T} \int\limits_{-T}^{T}\mathrm{d}t ~\text{tr}\left[ (e^{-i {\mathcal L}_0 t}\dyad{\chi_1^\prime}{\chi_2^\prime})^\dagger {\mathcal W}^\text{M} e^{-i {\mathcal L}_0 t}\dyad{\chi_1}{\chi_2} \right]\nonumber\\
&= \delta_{E_{\chi_1}{-}E_{\chi_2},E_{\chi^\prime_1}{-}E_{\chi^\prime_2}} \left( {\mathcal W}^\text{M}\right)_{\chi_2^\prime \chi_2}^{\chi_1^\prime \chi_1}.\label{eq:rotatingframe}
\end{align}
Hence, translated into the diagrammatic picture, all diagrams where the energy difference $E_{\chi_1^\prime}{-}E_{\chi_2^\prime}$ of the initial states is different from the energy difference $E_{\chi_1}{-}E_{\chi_2}$ of the final states are simply neglected. 
As a consequence, the off-diagonal elements of the matrix $\ga^\alpha_{\mu\nu}$ are set to zero.
This brings $\ga^\alpha_{\mu\nu}$ into a diagonal and positive semidefinite form. 
The secular approximation leads to the following two Lindblad operators 
\begin{align}
  &\ell^\alpha_{\sigma,1}=\sqrt{f_\alpha(\eps)}c^\alpha_{\sigma}(1-n_{\bar \sigma}), \nonumber \\
  &\ell^\alpha_{\sigma,2}=\sqrt{f_\alpha(\eps+U)}c^\alpha_{\sigma}n_{\bar \sigma}, \label{eq:secular}
\end{align} 
with coupling constants $\ga^\alpha_k=\Gamma$. Inserting the Lindblad operators into Eq. \eqref{eq:Lindbladform2}, we arrive at the master equation in the secular approximation $\dot \rho= \LS \rho$.
Interestingly, in contrast to the phenomenological model, we obtain two \textit{separate} Lindblad operators with $k=1,2$ for each $\alpha$ and $\si$ corresponding to the excitation of the first and the second electron, respectively. 
Nonetheless, for the limiting case $T\rightarrow0$ and $\eps < \eps_\text{F} < \eps + U$, the Lindblad operators agree with the phenomenological results of Eq. \eqref{eq:lcold} since one of the two Lindblad operators is always zero for a fixed $\alpha$ (the channels are open in only one direction). 
We infer that the coupling parameters are $\ga^+_\text{cold}=\ga^-_\text{cold}=\Gamma$ and, thus, they are equal for the in- and outgoing electrons. (Strictly speaking, this is only true for one and the same excitation energy. However, within the wide-band limit where $\Gamma$ is energy independent, it is generally true.) Inspecting also the limit of a hot bath $T\rightarrow \infty$, the Lindblad operators remain occupation dependent and, thus, deviate from our model of Eq. \eqref{eq:lhot}. In the next section, we discuss which is the correct $T\rightarrow \infty$ limit.
One major drawback of the secular approximation is the discontinuity when differences in the  excitation energies reach zero (cf. Kronecker delta in Eq. \eqref{eq:rotatingframe}), meaning that the terms are kept for vanishing differences while they are omitted for small differences. Thus, albeit guaranteeing positivity, it is a rather drastic approximation, because some diagrams involving coherences are completely omitted. Therefore, the secular approximation can lead in certain cases to unphysical consequences, as discussed in Refs. \onlinecite{SSM12,DTAVB18}.

\subsubsection{Coherent approximation} \label{subsubsec:coh}

\def\LC{{\cal L}^\text{C}}

In order to overcome the above mentioned difficulties, we propose a less radical approximation.
Instead of neglecting the arithmetic mean in the off-diagonal elements of $\ga^\alpha_{\mu\nu}$, we replace them by the geometric mean
\begin{align} \label{eq:approx}
 \Ga \,\frac{f_\alpha(\eps)+f_\alpha(\eps{+}U)}{2} \approx \Ga \sqrt{f_\alpha(\eps)f_\alpha(\eps{+}U)}. 
\end{align} 
Due to this replacement, the negative eigenvalue $\ga^\alpha_-$ (cf. Eq. \eqref{eq:eigenvalues}) is shifted to zero, whereas the trace of $\ga^\alpha_{\mu\nu}$ remains unchanged. Hence, positivity is reestablished in a minimal invasive way.
By performing a unitary transformation to diagonalize $\ga^\alpha_{\mu\nu}$ and rescaling the Lindblad operators, we arrive at
only one relevant Lindblad operator for each $\sigma$ and $\alpha$
\begin{align}
  \ell^\alpha_{\sigma}&=\ell^\alpha_{\sigma,1}+\ell^\alpha_{\sigma,2}\nonumber \\&=\sqrt{f_\alpha(\eps)}c^\alpha_{\sigma}(1-n_{\bar \sigma}) +\sqrt{f_\alpha(\eps+U)}c^\alpha_{\sigma}n_{\bar \sigma}, \label{eq:geometric}
\end{align}
with a rescaled $\ga^{\alpha}=\Gamma$. The resulting master equation will be abbreviated as $\dot \rho= \LC \rho$ denoting the \textit{coherent} approximation. 
In a sense, this approximation leads to a coherent combination of the two individual Lindblad operators from Eq. \eqref{eq:secular} that describe separately excitations of the first ($\ell^\alpha_{\sigma,1}$) and the second ($\ell^\alpha_{\sigma,2}$) electron. Put differently, whereas in the secular approximation we deal with an incoherent sum of those two processes, the coherent approximation generates a single coherent excitation.
Interestingly, the coherent approximation agrees with both our expectations for a hot bath (cf. Eq. \eqref{eq:lhot}) and for a cold bath (cf. Eq. \eqref{eq:lcold}). For the former case, we can infer that the coupling parameters are $\ga_\text{hot}^{\alpha}=\Gamma/2$, again independent of the direction $\alpha=\pm$. 
Moreover, it turns out that the finite-temperature Lindblad operator from Eq. \eqref{eq:geometric} can indeed be rewritten as an interpolation between the hot and cold limits and by comparison with Eq. \eqref{eq:lfinite} we find $g_\pm(T)=\pm \left[ \sqrt{f_\pm(\eps)}-\sqrt{f_\pm(\eps+U)} \right]$ as well as $h_+(T)=\sqrt{f_+(\eps+U)}$ and $h_-(T)=\sqrt{f_-(\eps)}$  satisfying $g_\alpha(T\rightarrow \infty)=0$ and $h_\alpha(T\rightarrow 0)=0$.  Thus, for one site $M=1$, the Liouvillians of the phenomenological approach and the coherent approximation are identical, ${\cal L}_\text{phen}=\LC$.
However, since the approximation of Eq. \eqref{eq:approx} is a rather ad-hoc replacement, we check, in the next step, whether the coherent approximation is physical and leads to better results than the secular approximation. 

Below, we compare three limiting cases, where we compare the coherent with the secular approximation:
\begin{enumerate}
\item In the limit $T\rightarrow 0$, the coherent approximation and the secular approximation coincide $\LC=\LS$ because the off--diagonal elements $\ga^\alpha_{12}=\ga^\alpha_{21}=\Ga \sqrt{f_\alpha(\eps)f_\alpha(\eps{+}U)}$ vanish. Thus, for small temperatures we expect the coherent approximation to be at least as good as the secular approximation.

\item In the limit $T\rightarrow \infty$, the function $\ga_{\alpha,\si}^\text{(1)}(\tau)$ becomes delta peaked in $\tau$ so that the width of the first-order diagrams becomes zero. Then, we expect the Markovian assumption ${\cal W}^\text{M}$ to be exact. The arithmetic and the geometric mean give rise to identical off--diagonal elements $\ga^\alpha_{12}=\ga^\alpha_{21}=\Gamma/2$ meaning Eq. \eqref{eq:approx} is an identity and, hence, the coherent approximation becomes also exact. In contrast, the secular approximation sets the off-diagonal elements to zero $\ga^\alpha_{12}=\ga^\alpha_{21}=0$. Thus, for high temperatures, the coherent approximation is evidently better than the secular approximation.

\item When differences in the excitation energies are small (to be specific: when $\eps+U-\eps=U$ is small), we find that the coherent approximation of Eq. \eqref{eq:approx} is an equality up to the first order in $U$. Hence, for vanishing interaction $U=0$, Eq. \eqref{eq:approx} again is exact and the off-diagonal elements remain unchanged $\ga^\alpha_{12}=\ga^\alpha_{21}=\Ga f_\alpha(\eps)$ (as it would be in the secular limit for $U=0$). But, now, instead of an abrupt change (as in the secular approximation), the off-diagonal elements are also kept for small values of $U$, leading to a continuous transition. In contrast, for strong interactions $U\gg k_\text{B}T$ in the particle-hole symmetric case (where the excitation energies $\eps$ and $\eps +U$ are centered symmetrically around the Fermi energy $\eps_\text{F}$), the geometric mean $\Ga\sqrt{f_\alpha(\eps)f_\alpha(\eps{+}U)}$ vanishes exponentially with $U$. Thus, in the limit $U\rightarrow \infty$ the coherent and the secular approximation coincide.

\end{enumerate}

To conclude, we claim that the coherent approximation of Eq. \eqref{eq:approx} is physically justified and less radical than the secular approximation. Most importantly, it also guarantees positivity. 

\subsubsection{Modified Rules} \label{subsubsec:mod}

Instead of performing the coherent approximation in the way we presented, we can also embed the modifications on the level of the diagrammatic rules from equation \eqref{eq:diagramstolindblads}. There, every real part of a diagram contributes to the transition rate with a term $\sim f_\alpha(\Delta E)$ where the energy difference $\Delta E$ is always determined by the states directly before and after the right vertex (right green dot). Now, by replacing $f_\alpha(\Delta E)\rightarrow \sqrt{f_\alpha(\Delta E)f_\alpha(\Delta E^\prime)} $, where $\Delta E^\prime$ is the energy difference at the left vertex (left green dot), we arrive in an alternative way at the coherent approximation. Doing so, we can conveniently generalize the approximation to other systems. Note that our coherent approximation is in accordance with a phenomenological approach proposed in Ref. \onlinecite{KFW18} which was discussed further in Ref. \onlinecite{PE19}. 

\subsection{$M$ sites}

It turns out that for a single--site Hubbard model, differences between the secular and coherent approximation cannot be observed, because they differ only in transitions between coherences of the form $\dyad{0}{\sigma}\leftrightarrow\dyad{\bar{\sigma}}{\uparrow\downarrow}$. 
Since those coherences are between states of different Fermion parity, they are irrelevant for any measurable quantity, see fermion-parity superselection postulate in Refs. \onlinecite{WWW52,SSHSW16}.
However, for a Hubbard model with two sites, respective transitions $\dyad{0,\uparrow\downarrow}{\sigma,\bar{\sigma}}\leftrightarrow\dyad{\bar{\sigma},\uparrow\downarrow}{\uparrow\downarrow,\bar{\sigma}}$ now become relevant. This motivates us to look closer at the dynamics of a Hubbard dimer (see Sec. \ref{sec:dimer}). To do so, we have to extend the formalism to larger systems with $M$ sites.

\subsubsection{Non-Local Lindblad Operators}

We discuss both the secular and coherent approximation for an arbitrary number of sites $M$. Note that for $M \geq 2$ the tunneling amplitude $J$ between the sites complicates the spectrum of excitation energies $\Delta E$.
\paragraph{Secular approximation:} By using equation  \eqref{eq:rotatingframe} we can apply the secular approximation straightforwardly  to larger systems, again, by discarding all those diagrams where the energy difference of the initial states differs from the energy difference of the final states. Then we arrive at
\begin{align}
\ell^+_{m,\sigma}(\Delta E)&=\sum_{\chi,\chi^\prime}\delta_{\Delta E,E_\chi{-}E_{\chi^\prime}}\sqrt{f_+(E_\chi{-}E_{\chi^\prime})} \mel{\chi}{c\s_{m,\si}}{\chi^\prime} \dyad{\chi}{\chi^\prime},  \nonumber \\
\ell^-_{m,\sigma}(\Delta E)&=\sum_{\chi,\chi^\prime}\delta_{\Delta E,E_{\chi^\prime}{-}E_{\chi}}\sqrt{f_-(E_{\chi^\prime}{-}E_\chi)} \mel{\chi}{c_{m,\si}}{\chi^\prime} \dyad{\chi}{\chi^\prime},\label{eq:secgen}
\end{align}
which describe separate transitions for each (positive) single-electron excitation energy $\Delta E$. For $M=1$ we arrive at the Lindblad operators of equation \eqref{eq:secular}.  Again, we abbreviate the corresponding Liouvillian with $\LS$ where $\text{S}$ is short for the secular approximation.
\paragraph{Coherent approximation:} For the coherent approximation, on the other hand, a procedure as in Sec. \ref{subsubsec:coh} becomes unhandy. However, a straightforward extension to larger systems can be performed by using the modified diagrammatic rules of Sec. \ref{subsubsec:mod}. Then, we arrive at the following Lindblad operators
\begin{align}
\ell^+_{m,\sigma}&=\sum_{\chi,\chi^\prime}\sqrt{f_+(E_\chi{-}E_{\chi^\prime})} \mel{\chi}{c\s_{m,\si}}{\chi^\prime} \dyad{\chi}{\chi^\prime},  \nonumber \\
\ell^-_{m,\sigma}&=\sum_{\chi,\chi^\prime}\sqrt{f_-(E_{\chi^\prime}{-}E_\chi)} \mel{\chi}{c_{m,\si}}{\chi^\prime} \dyad{\chi}{\chi^\prime},\label{eq:cohgen}
\end{align} 
which reduce for a single site to the known results of the coherent approximation from Eq. \eqref{eq:geometric}. Again, as in Eq. \eqref{eq:geometric}, we find that the Lindblad operators are a coherent sum of those found in the secular approximation, $\ell^\alpha_{m,\si}=\sum_{\Delta E} \ell^\alpha_{m,\si}(\Delta E)$.  We will denote the corresponding Liouvillian, as before, $\LC$ where $\text{C}$ is short for the coherent approximation.

Interestingly, although we model local baths only, we find that for $\LS$ and $\LC$ with $M\geq 2$ the Lindblad operators include non-local transitions in space that cause changes at sites $m^\prime\neq m$.  Formally, this non-locality originates in the integration over the past time in Eq. \eqref{eq:transitionrate}, which in the Markov assumption $t_0\rightarrow -\infty$ leads to non-local transitions even for early times $t-t_0 \lesssim \tau_\text{c}$. Hence, the Lindblad operators $\ell^\alpha_{m,\sigma}$ as well as $\ell^\alpha_{m,\sigma}(\Delta E)$ describe effective transitions where not only an electron at site $m$ with spin $\si$ enters or leaves, but at the same time $J$-induced inter-site tunneling events can happen. Consequently, the Lindblad operators take for larger systems a more involved form.

\subsubsection{Local Lindblad Operators}

Next, we want to discuss a less sophisticated but inherently local approach to model the coupling to the baths. Its main benefit is its simplicity and the ability to directly apply the Lindblad operators we have obtained for a single site. An obvious (but not fully correct) way to describe a situation where every site is coupled to a local bath is by simply promoting the Lindblad operators for one site with a site index $\ell^\alpha_\sigma \rightarrow \ell^\alpha_{m,\sigma}$. Then, all dissipative effects are assumed to be independent of the inter-site tunneling amplitude $J$. We will call this assumption the local approximation and denote the respective Liouvillians as  $\LS_{\text{loc}}$ and $\LC_{\text{loc}}$ for the secular approximation \eqref{eq:secular} and the coherent approximation \eqref{eq:geometric}, respectively. In contrast to the full Liouvillians $\LS$ and $\LC$, here, the Lindblad operators induce only local transitions. Interestingly, we find also for multiple sites $M\geq 2$ that the phenomenological approach (cf. Eqs. \eqref{eq:lhot}-\eqref{eq:lfinite}) is identical to the local coherent approximation, ${\cal L}_\text{phen}=\LC_\text{loc}$. Note that we may obtain the local approximations $\LS_{\text{loc}}$ and $\LC_{\text{loc}}$ by setting $J=0$ in the Lindblad operators of Eq. \eqref{eq:secgen} and Eq. \eqref{eq:cohgen}, respectively. 

The local approximation seems reasonable as long as we are in the limit of a weak hopping $J \sim \Gamma$ such that the corrections to the excitation energies $\eps$ and $\eps +U$ are negligible within the leading order perturbation theory in $\Gamma$. However, this argument may not hold for times $t\gg \Ga^{-1}$ since small errors may accumulate over time. Moreover, in Ref. \onlinecite{LK14}, it is stated that the local approach may even lead to a violation of the second law of thermodynamics. 

Interestingly, for the coherent approximation, we find for a hot bath $T\rightarrow \infty$  that local and non-local Liouvillians are identical, $\LC=\LC_\text{loc}$, and give both exact results (since the replacement $f_\alpha(\Delta E)\rightarrow \sqrt{f_\alpha(\Delta E)f_\alpha(\Delta E^\prime)} $ from Sec. \ref{subsubsec:mod} becomes an identity). Hence, for sufficiently high temperatures $k_\text{B}T\gg J,U$ the non-local effects are negligible and the local coherent approximation $\LC_\text{loc}$ should perform very well. For smaller temperatures, however, we have to be careful with the local approximation and study explicitly the consequences of non-local transitions.

\section{Hubbard dimer}\label{sec:dimer}

Below, we will be interested in a Hubbard dimer, which is the smallest possible system where all energy scales, i.e., the inter-site tunneling amplitude $J$, the on-site Coulomb interaction $U$, the system-bath tunnel-coupling strength $\Ga$ and the temperature $T$ of the baths, appear.
The Hamiltonian takes the form $H_\text{s}=H_J+H_U+H_\eps $ with $H_J=-J \sum_{\sigma} \h c\s_{1,\sigma} \h c_{2,\sigma} +\text{h.c.}$ for tunneling between the two sites, 
$H_U= \sum_m U \h n_{m,\sd} \h n_{m,\su} $ for the on--site Coulomb interaction and $H_\eps=\eps \sum_{m,\sigma}  \h n_{m,\sigma}$ for the single-electron energies, as introduced in Eq. \eqref{eq:Hamiltonian}.  The system can be experimentally realized as a double quantum dot coupled to (separate) electronic leads \cite{PSSetal12,MBHetal12,WPDetal13}.
First, we discuss the dynamics of simple observables as the total particle number and the total spin and discuss the characteristic decay. Then, we focus on the limit of hot baths, $T\rightarrow \infty$, which is suited very well to estimate the validity of both the local and non-local secular approximations $\LS_\text{loc}$ and $\LS$, because for hot baths the coherent approximations $\LC_\text{loc}=\LC$ give exact results. Next, we elaborate the limit of cold baths, $T\rightarrow 0$, where both local approximations yield identical results $\LS_\text{loc}=\LC_\text{loc}$. We compare the local as well as the full Liouvillians $\LC$ and $\LS$ with a full solution to the time-dependent equation \eqref{eq:timelocal}.
At last, we mention shortly the case of small but finite temperature.

\subsection{Dynamics of Simple Observables}\label{sec:observables}

To study time-dependent observables $O(t)$, it is useful to work in the Heisenberg picture by employing the super-adjoint Lindblad equation
\begin{align} \label{eq:adj-Lindblad}
  & \dot{ O}(t) = \mathcal{L}^{ \dagger} \h O =i\, [H_\text{s},O] \nonumber \\ 
  &+ \Ga \sum_{m,\alpha,\sigma,k} \left[\h \ell^{\alpha\dagger}_{m,\sigma,k}\, \h O\, \h \ell^\alpha_{m,\sigma,k} - \frac1{2}\{\h \ell^{\alpha\dagger}_{m,\sigma,k}\,\h \ell^\alpha_{m,\sigma,k}, O\} \right],
\end{align}
with $\L\s$ being the adjoint Liouvillian of either one of the four possible $\LS,\LC,\LS_\text{loc}$ or $\LC_\text{loc}$. At first, we discuss the dynamics of basic observables as the $z$-component of the total spin $S_z(t)=\sum_m (n_{m,\su}-n_{m,\sd})/2$ and the total occupation number $N(t)=\sum_{m,\sigma} n_{m,\sigma}$. For those observables, we find $\left(\LS_\text{loc}-\LC_\text{loc}\right)^\dagger O(t)=0$ and  $\left(\LS-\LC\right)^\dagger O(t)=0$ with either $O(t)=S_z(t)$ or $O(t)=N(t)$, meaning secular and coherent approximation give the same results. For the local limit $\LS_\text{loc}$ and $\LC_\text{loc}$ the super--adjoint Lindblad Eq. \eqref{eq:adj-Lindblad} simplifies to a closed ordinary differential equation and
we find (in agreement with Ref. \onlinecite{SGKB10}) for the total spin $S_z$
\begin{align}
  \dot{ S}_z(t) = \lambda_{S_z} S_z(t), \label{eq:spinz}
\end{align}
with $\lambda_{S_z} = -\Ga\left[1 + f(\eps+U) - f(\eps)\right]$ 
and for the total occupation number
\begin{align}
  \dot{ N}(t) = \lambda_N \left[ N(t)- N(\infty) \right], \label{eq:number}
\end{align}
where $\lambda_N =-\Ga\left[1 - f(\eps+U) + f(\eps)\right]$ and $N(\infty)={-}2 n_s f(\eps)\Ga/\lambda_N$ with $n_s$ being the number of sites. These results are also valid for an arbitrary number of sites.
The equations have exponentially decaying solutions, $O(t) = O(\infty) + \left[ O(0) - O(\infty) \right]\, e^{\lambda\, t}$ with the exponents given by $\lambda_{S_z}$ and $\lambda_N$, respectively, and the asymptotic values $O(\infty)$ being zero for the spin and $N(\infty)$ for the occupation number.

For the non-local methods $\LS$ and $\LC$ the decay of $N$ and $S_z$ becomes more involved, and we do not obtain a simple closed differential equation. However, since the only possible relaxation rates that can appear are the eigenvalues $\lambda_j$ of the respective Liouvillian ${\L}$, it is useful to perform a spectral decomposition of the expectation values
\begin{align}
\ev{O}(t)=\left(O,\rho(t)\right)=\sum_j e^{\lambda_j t}(O,r_j)(l_j,\rho_0),\label{eq:spectral}
\end{align}
with left and right eigenvectors of $\L$ defined via ${\L} r_j = \lambda_j r_j$ and ${\L}^\dagger l_j=\lambda_j l_j$. 
They are orthogonal $(l_i,r_j)=\delta_{ij}$ and fulfill the completeness relation $\sum_j r_j l_j^\dagger=\1$. Doing so, we find analytical results for the non-local methods in the limiting cases $T\rightarrow \infty$ and $T\rightarrow 0$ with $\eps < \eps_\text{F} < \eps + U$. For comparison we summarize the relevant decay constants with the smallest nonzero absolute values
\begin{align}
  \begin{array}{|c|c|c|} \hline
    & \lambda_{N} & \lambda_{S_z} \\ \hline
      \text{local } T\rightarrow \infty & -\Ga & -\Ga \\ \hline
         \text{local } T\rightarrow 0 & -2\Ga & 0 \\ \hline
            \text{non-local } T\rightarrow \infty & -\Ga & -\Ga \\ \hline
   \text{non-local } T\rightarrow 0 & -2\Ga(1\pm\delta) & -2\Ga(1\pm\delta) \\ \hline
  \end{array}
\end{align}
where $\delta=J/\sqrt{(4J)^2+U^2}$. It is interesting to observe that according to the local approaches the relaxation of the total spin ${S_z}$ becomes slower with decreasing bath temperature until a total freeze at $T\rightarrow 0$ while the relaxation of the occupation number $N$ becomes faster at lower temperatures.
The latter is an interaction effect and can be explained by the increasing asymmetry between the bath occupations at different energies, $f(\eps)$ and $f(\eps+U)$, which at lower temperatures increase the couplings for the uni--directional processes \textit{double occupation} \tra \textit{single occupation} and \textit{empty site} \tra \textit{single occupation}, leading to a faster relaxation of $N$ to the one--electron--per--site equilibrium.
On the other hand, for the spin relaxation a multiple exchange of electrons between the sites and the baths is needed since in leading order in $\Ga$ a direct spin relaxation is suppressed within the local approach \cite{CSGKB12}.
So when the electron transfer saturates quickly (for low temperatures) the spin decay proceeds very slowly.

By employing the full non-local Liouvillians $\LS$ or $\LC$ we find for the limit of hot baths $T\rightarrow \infty$ identical results, but for cold baths $T\rightarrow 0$ there are strong deviations, namely the spin $S_z$ is actually not conserved but decays in a more complicated fashion with characteristic decay constants $\propto -2\Ga(1 \pm \delta)$. This decay originates from effective processes where, e.g., the electron with spin $\si$ tunnels from one site $m$ to a neighboring site $m^\prime$ ($\propto J$) and at the same time an electron with opposite spin $\bar{\si}$ enters from the bath at site $m$ ($\propto \Ga$) resulting in a  mechanism for direct spin relaxation. Hence, for the dynamics of $S_z$, we find a drastic discrepancy where non-local transitions become very important for small temperatures. In contrast, for $N$, the discrepancy is rather a small correction in the decay rate if the inter-site tunneling amplitude $J$ is small compared with $U$. In consequence, when discussing small temperatures, a comparison with models beyond the local approximation becomes necessary. 

\subsection{Hot baths}

\begin{figure}[ht]
  \includegraphics[width=0.9\linewidth]{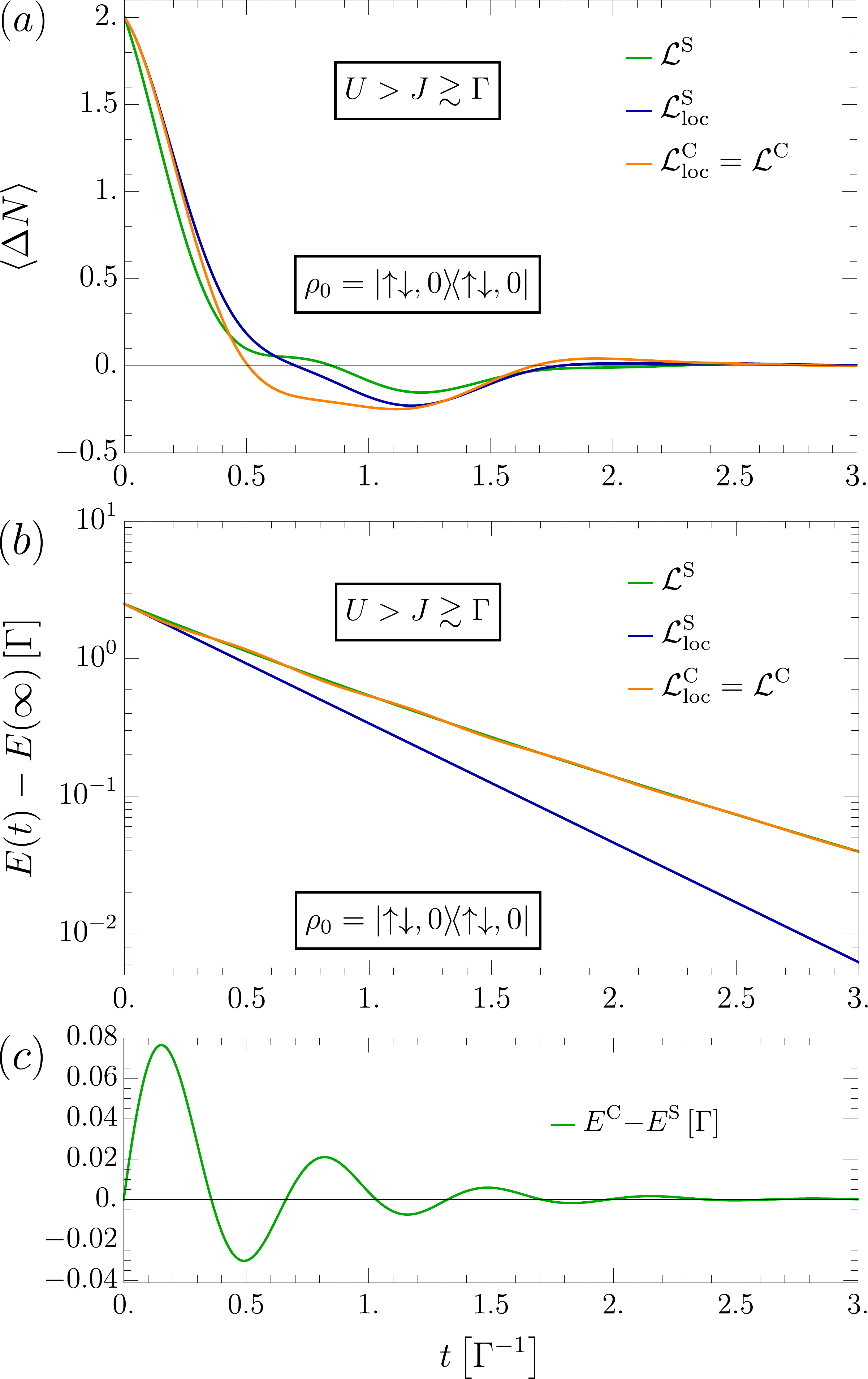}
  \caption{(a) Particle asymmetry $\ev{\Delta N}$ and (b) energy $E=\ev{H_\text{s}}$ as function of time starting from an initial state $\rho_0=\dyad{\su,0}$. We compare the local secular approximation $\LS_\text{loc}$ (blue line) and the non-local secular $\LS$ approximation (green line) with the exact results given by the local coherent approximation $\LC_\text{loc}=\LC$ (orange line). In (c) we resolve the differences $E^\text{C}{-}E^\text{S}$ between the non-local secular approximation $\LS$ and the exact results $\LC$. The parameters are $U=5\Ga$, $J=2\Ga$ and $T\rightarrow \infty$.}
  \label{fig:diffnumbers}
\end{figure}

Remember that for hot baths (formally at infinite temperature) the local coherent approximation $\LC_\text{loc}$ is exact (and equal to $\LC$) and hence we can determine how far off the local and non-local secular approximation $\LS_\text{loc}$ and $\LS$ is, respectively.  In all cases, the stationary state of the system becomes fully mixed 
\begin{equation}
  \h\rho_{\text{st}} = \frac1{16} \sum_{\chi_1, \chi_2 \in \{0, \su, \sd, \su\sd\} } \dyad{\chi_1, \chi_2}{\chi_1, \chi_2},
\end{equation}
containing all possible occupations $\ket{\chi_1,\chi_2}$ at both sites (first and second entry) coming with equal probabilities. Hence, the stationary state agrees with the Gibbs state $\rho_\text{G}=e^{-\beta (H_\text{s}-\eps_\text{F} N)}/Z $ for $\beta\rightarrow 0$.
Recalling Sec. \ref{subsec:onesite}, we find that differences between the local secular ($\LS_\text{loc}$) and the local coherent ($\LC_\text{loc}$) approximation scale with $\propto \sqrt{f_\alpha(\eps)f_\alpha(\eps+U)}$, which for the particle-hole symmetric case $f_\alpha(\eps+U)=1-f_\alpha(\eps)$ (shown in Fig. \ref{fig:bath}) can be rewritten as $\propto \lbrack 2 \cosh(\beta U/4)\rbrack^{-1}$. This expression diminishes exponentially for high values of $\beta U \gg 1$. 
In contrast, for high temperatures (small $\beta$) and moderate interactions $U$, we should observe differences in the dynamics. In Fig. \ref{fig:diffnumbers}, we show the time-dependent expectation value of $\Delta N=n_1{-}n_2$ with occupation number operators $n_1$ and $n_2$ as a function of time for both the local and the non-local secular approximation against the exact coherent approximation. 
The system is initially prepared in the state $\rho_0=\dyad{\uparrow\downarrow,0}$ and interacts with a hot bath ($T\ra\infty$) at each site. 
Despite a qualitative agreement, we can see some differences in the time evolution. 
The differences originate in those processes involving the transitions $\dyad{\uparrow\downarrow,0}\rightarrow \dyad{\uparrow\downarrow,0}{\bar{\sigma},\sigma}\rightarrow\dyad{\uparrow\downarrow,\bar{\sigma}}{\bar{\sigma},\uparrow\downarrow}$. The first step is due to a coherent evolution generated by $H_J$ and the second step corresponds to an electron entering from the bath at site $2$ with spin $\bar{\sigma}$. 
While those processes are completely neglected in the secular approximation, which does not couple coherences with different energies (cf. Eq. \eqref{eq:rotatingframe}), the coherent approximation takes them correctly into account.
Furthermore, if we look at the characteristic energy decay $E=\ev{H_\text{s}}$ (and subtract the stationary value $E(\infty)=2\eps$), we find different relaxation rates for the respective approximations. For the local secular approximation, the energy relaxation goes simply with $-2\Ga$, whereas for the coherent approximation we find a decay with $-2\Ga< \Re\la_E < -\Ga$, for instance $\Re\lambda_E\approx -1.15 \Gamma$ for $U=5\Ga, J=2\Ga$, nearly half as slow. The value can be determined analytically as
\begin{align}
  & \lambda_E = {-}2\Ga{+} \nonumber  \\ 
  & \frac{1}{\sqrt{2}} \sqrt{\Ga^2 - U^2 - (4J)^2 {+} \sqrt{[\Ga^2 - U^2 - (4J)^2]^2 + 2 (4J)^2 \Ga^2}},
\end{align}
which has a non-vanishing oscillatory part $\Im \lambda_E$ as well as a decaying part $\Re \lambda_E$ that gives $-2\Gamma$ in the limit of either $U\rightarrow \infty$ or $J\rightarrow 0$
and $-\Gamma$ for $U, J \ll \Ga$. Surprisingly, the non-local secular approximation $\LS$ gives the same real part of the relaxation rate $\lambda_E$, but misses the imaginary part and, thus, does not describe the oscillatory part at all, see Fig. \ref{fig:diffnumbers}c. To conclude, for the Hubbard dimer, there are qualitative differences in the relaxation dynamics, because the secular approximation, indeed, misses systematically some relevant processes. 

\subsection{Cold baths}

We consider the case of cold baths (formally at zero temperature) with  $\eps < \eps_\text{F} < \eps + U$ where the local coherent and the local secular approximations coincide, $\LS_\text{loc}=\LC_\text{loc}$, and we explicitly discuss the importance of non-local effects that are included in $\LS$ and $\LC$. 

The stationary state $\rho_\text{st}$ depends strongly on the inclusion of non-local processes. In the local limit, we find that $\rho_\text{st}$ belongs to a degenerate subspace spanned by the spin triplet states 
\begin{align} \label{eq:triplet}
  \ket{T_{-}} &= \ket{\sd,\sd}, & \ket{T_{+}} &= \ket{\su,\su}, & \ket{T_0} &= \frac{1}{\sqrt{2}}(\ket{\sd,\su} + \ket{\su,\sd}),
\end{align}
which fulfill $H_J \ket{T_l}=H_U \ket{T_l}=0$ and $\vb{S}^2 \ket{T_l}=2 \ket{T_l}$ with $l\in\{-,0,+\}$. The total spin is defined as $\vb{S}=\sum_{m,\si,\si^\prime}c^\dagger_{m,\si}\boldsymbol{\tau}_{\si\si^\prime}c_{m,\si^\prime}/2$ with Pauli matrices $\boldsymbol{\tau}=(\tau_x,\tau_y,\tau_z)$ such that $S(S+1)$ are the eigenvalues of $\vb{S}^2$ with $S=1$ for the triplets.
However, the eigenstate of  the grand-canonical Hamiltonian $H_\text{s}-\eps_\text{F}N$ with the lowest eigenvalue is not a triplet state, but rather a modified singlet state of the form $\ket{\tilde{S}}=(\ket{S}+\kappa\ket{DH})/\sqrt{1+\kappa^2}$ with $\ket{S}=\frac{1}{\sqrt{2}}(\ket{\sd,\su} - \ket{\su,\sd})$ being the actual singlet and $\ket{DH}=(\ket{\uparrow\downarrow,0}+\ket{0,\uparrow\downarrow})/\sqrt{2}$ being a symmetric superposition of doublon-holon states. The coefficient is given by $\kappa=(4J/U)/\lbrack 1+\sqrt{1+(4J/U)^2} \rbrack$ and diminishes for small ratios $J/U$. Apparently, it is not possible to arrive at the lowest-energy state $\ket{\tilde{S}}$. The reason is, that within the local approach a double occupation is unstable and must always decay. Consequently, $\ket{\tilde{S}}$ cannot become stationary.  

Employing the Liouvillians $\LS$ and $\LC$, on the other hand, we find that if non-local transitions are included, the subspace for the stationary state $\rho_\text{st}$ is extended by exactly $\ket{\tilde{S}}$. The non-local transitions are such that, now, also a double occupation of one site can be energetically favorable. Only then, the findings are consistent with the Gibbs state. However, note that in the limit $T\rightarrow 0$, the stationary state is actually not unique, because the system can be in any linear combination of the four states $\ket{S}$ and $\ket{T_l}$ as well as in mixture of them. Hence, in Liouville space we have a sixteen-dimensional degenerate subspace to the eigenvalue $\Re \lambda_0=0$ of $\LS$ (or $\LC$). Interestingly, if a certain initial state $\rho_0$ is given, the stationary solution is completely determined by sixteen conserved quantities of the system, which are given by the left eigenvectors with eigenvalue  $\Re \lambda_0=0$ of the Liouvillian \cite{AJ14}. In contrast, for finite temperatures there is only one conserved quantity, namely probability, and thereby only one unique stationary state.

Next, we discuss two specific examples of a quantum quench to highlight the main qualitative differences in the dynamics between the local ($\LS_\text{loc}=\LC_\text{loc}$) and non-local ($\LS$ and $\LC$) methods. Furthermore, we consider also the full solution to the raw equation \eqref{eq:timelocal}, where the Markovian limit $t_0\rightarrow \infty$ has not been applied yet and the transition rates from Eq. \eqref{eq:transitionrate} still depend on the time that has elapsed since $t_0$ (again we consider only the real part of the diagrams). Comparing with this brute force approach, we can better estimate the validity of the different approximations.

\begin{example} 

The system is initially prepared in the state where a spin-$\uparrow$ electron occupies one site $\rho_0=\dyad{\su,0}$ (with $U = 10\, \Ga, J=2\, \Ga$). Such a state can be experimentally prepared by means of a magnetic field that favors the spin-$\su$ direction and by a specific tuning of the excitation energies via gate voltages.  
By quickly resetting the parameters, one can study the quench dynamics. 
Once in contact with the cold baths, the system relaxes quickly to a state in which each site is occupied by exactly one electron. We find, that the number asymmetry $\Delta N=n_1-n_2$ decays at a rate $\approx -2\,\Ga$ for $U\gg \Gamma,J$, thus, leading to a quick end of the coherent oscillations of particles, see Fig. \ref{fig:spinupdecay}a. For $\Delta N$, all methods agree very well with each other.
For the total spin $z$-component $S_z$, we found that in the local approaches we have $\dot{S}_z=0$, whereas when non-local transitions are considered we have $\dot{S}_z\neq 0$. For the spin asymmetry between both sites $\Delta S_z=S_{z,1}-S_{z,2}$, the differences in the dynamics are even more intriguing, see Fig. \ref{fig:spinupdecay}b. For short times, the decay is fast with a rate $-2\Ga$, which is similar for all approaches.  However, for longer times, we observe either slow decaying oscillations (blue line) predicted by the local approaches, fast decaying oscillations (green line) predicted by the non-local secular approximation, or no decay at all (black and orange line) predicted by the non-local coherent approximation and the full method from  equation \eqref{eq:timelocal}, respectively. In the local approaches, the slow decay of $\Delta S_z$ is determined by the rate $\approx -8J^2\Gamma/U^2 $ for $U\gg \Gamma,J$.
This slow relaxation of spin oscillations originates in the decay of coherences of the form $\dyad{S}{T_l}$. It turns out that the singlet state $\ket{S}$ has no direct decay possibility in the local approach, as it can only decay via processes involving a prior coherent evolution to the double occupation of one site (suppressed by $U$) and a successive exchange of electrons with the baths, leading slowly to a triplet state, a dead--end of the evolution. In contrast, for the non-local approaches we have either a fast decay with $\approx -2\Ga$ for the secular approximation $\LS$ (green line) or no decay at all for the coherent approximation $\LC$ (black line). The latter case is indeed possible, since the coherences $\dyad{T_l}{\tilde{S}}$ that lead to spin oscillations (but no charge oscillations) fulfill $\LC \dyad{T_l}{\tilde{S}}=-i\lbrack \sqrt{(4J)^2{+}U^2}{-}U\rbrack /2 \dyad{T_l}{\tilde{S}}$. Thus, they are eigenvectors of $\LC$ to an eigenvalue with zero real part (no decay) but finite imaginary part (coherent oscillation). In other words, the spin oscillations happen in a decoherence-free subspace of dark states \cite{BP12,BTJ19}. To judge which method gives the most accurate results, we compare with the full method from equation \eqref{eq:timelocal} (orange line). It also shows forever-lasting coherent spin oscillations and apart from small deviations, it agrees very well with the coherent approximation $\LC$. 
\begin{figure}[ht]
 \includegraphics[width=0.9\linewidth]{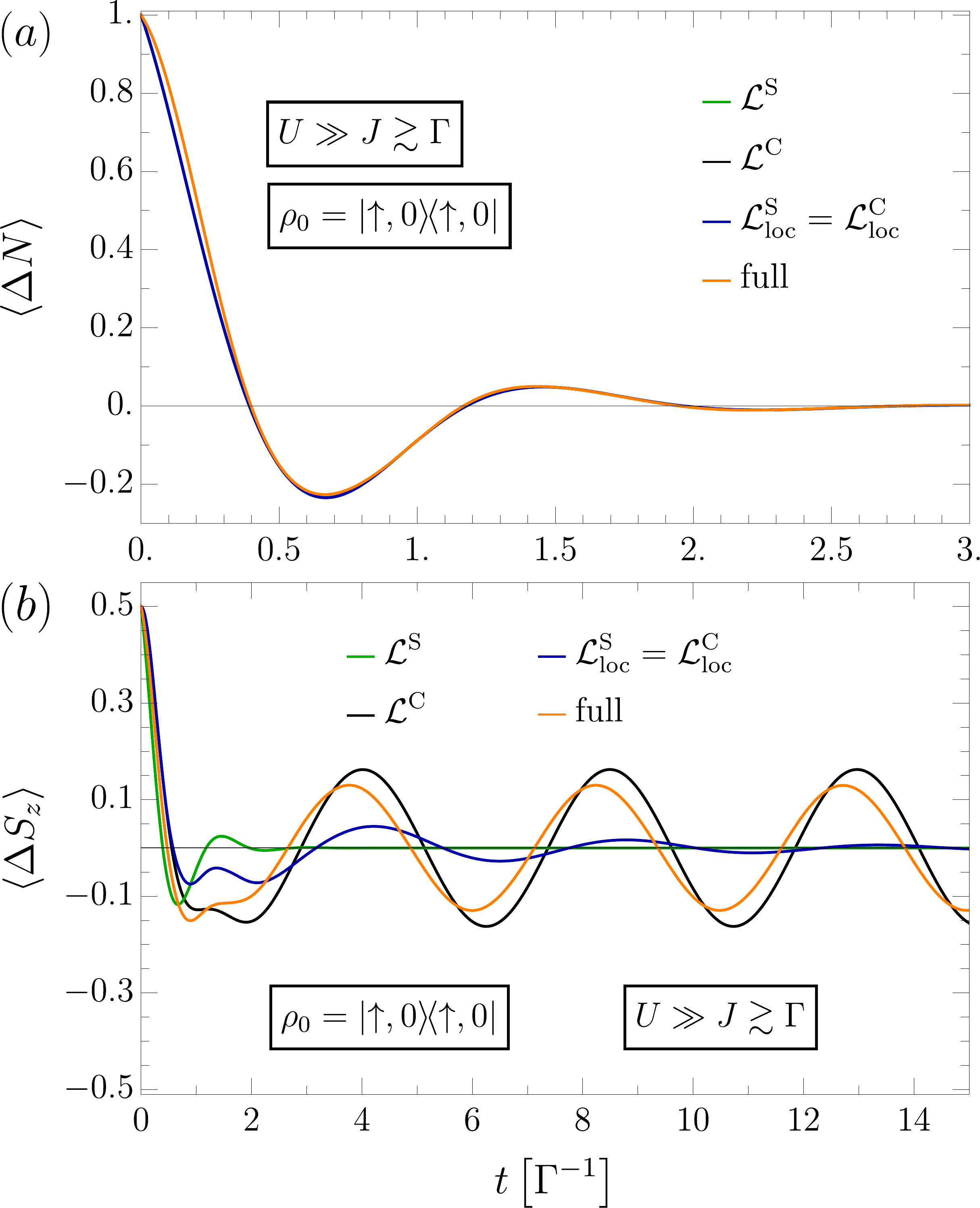}
  \caption{ (a) Particle asymmetry $\ev{\Delta N}$ and (b) spin asymmetry $\ev{\Delta S_z}$ as functions of time starting from an initial state $\rho_0=\dyad{\su,0}$. We compare the local methods $\LS_\text{loc}=\LC_\text{loc}$ (blue line) with the non-local secular $\LS$ (green line) and the non-local coherent approximation $\LC$ (black line). Moreover, we show the full solution to the time-dependent problem from Eq. \eqref{eq:timelocal} (orange line). The parameters are $U=10\Ga$, $J=2\Ga$ and $T\rightarrow 0$.}
  \label{fig:spinupdecay}
\end{figure}
\end{example}

\begin{figure}[ht!]
	\includegraphics[width=0.9\linewidth]{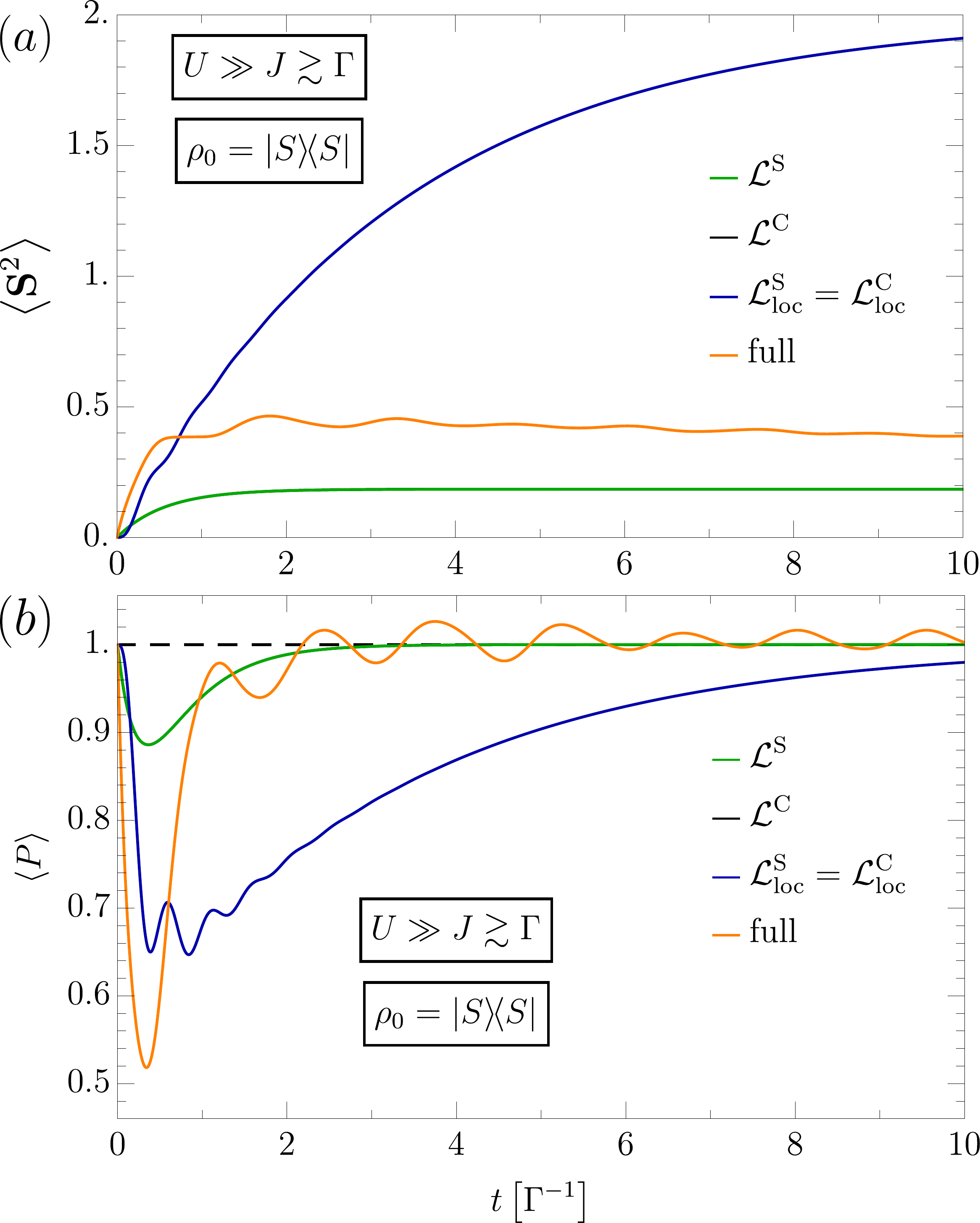}	
  \caption{(a) Spin squared $\vb{S}^2$ and (b) parity $P=(-1)^N$ as functions of time starting from an initial state $\rho_0=\dyad{S}$. We compare the local methods $\LS_\text{loc}=\LC_\text{loc}$ (blue line) with the non-local secular $\LS$ (green line) and the non-local coherent approximation $\LC$ (black line). The latter two methods (green and black line) agree for the shown observables. Moreover, we show the full solution to the time-dependent problem from Eq. \eqref{eq:timelocal} (orange line). The parameters are $U=10\Ga$, $J=2\Ga$ and $T\rightarrow 0$.}
  \label{fig:sixobs}
\end{figure}

\begin{example}
The system is initially prepared in a singlet state $\h \rho_0 = \dyad{S}$ (with $U = 10\, \Ga, J=2\, \Ga$). In Fig. \ref{fig:sixobs}a-b, we show the expectation values of the total spin squared $\vb{S}^2$  as well as the Fermion parity operator $P=(-1)^N$ as functions of time with the initial values $S(S+1)=0$ and $(-1)^N=1$, respectively. 
For the local methods (blue line), the spin squared $\vb{S}^2$ increases gradually from the initial value to the one of the triplet states with $S(S+1)=2$. Again, the relaxation proceeds slowly, because the singlet can only decay via a prior coherent oscillation to $\ket{DH}$ which is strongly suppressed for $U\gg J$. The respective decay constant is now $\approx -12 J^2\Gamma/U^2$ for $U\gg \Gamma,J$. The non-local approaches (green line for $\LS$ and black line for $\LC$) give identical results. The spin squared saturates quickly with $-2\Ga(1-\delta)$ to a final value that is much lower than the value expected for triplet states. The reason is that the singlet state $\ket{S}$ has a big overlap $\braket{\tilde{S}}{S}$ with the additional stationary state $\ket{\tilde{S}}$ that is missing in the local approach. Thus, triplet states contribute only partially to the stationary value and since $\ket{\tilde{S}}$ has zero spin, the final value of $\vb{S}^2$ is much lower. Interestingly, we find for the full model (orange line) that for small times the curve agrees more with the local approach while for larger times, albeit the presence of oscillations, the curve behaves more similar to the non-local methods. Consequently, due to the fast increase in the beginning the stationary value must lie higher. By considering the integrated relaxation rates $\int_0^{t{-}t_0}\mathrm{d}\tau \gamma_\pm(\tau)$ from Eq. \eqref{eq:transitionrate}, indeed, we find that for short times $t{-}t_0 < \hbar/\vert \Delta E{-}\eps_\text{F}\vert_\text{max}$ they describe only local transitions, what agrees with physical intuition, however, imitating a situation where the dimer is virtually coupled to hot baths (instead of cold ones). Only after a time $t{-}t_0> \tau_\text{c}=\hbar/\vert \Delta E{-}\eps_\text{F}\vert_\text{min}$ the rates saturate and effective non-local transitions are allowed, which results in a better agreement with the non-local methods $\LS$ and $\LC$. Considering the time evolution of the parity of the state, we find for all methods roughly similar trends. Starting from an initial value $+1$, in all cases the singlet state $\ket{S}$ is not stationary and due to intermediate transitions to odd parity states with one or three electrons the parity first decreases towards $-1$ and then finally reaches the stationary value $+1$ again. As before, for the local methods (green line) the parity reaches asymptotically $+1$ with the rate $\approx -12 J^2\Gamma/U^2$  for $U\gg \Gamma,J$. The non-local approaches (green line for $\LS$ and black line for $\LC$) give again identical results, and the asymptotic value $+1$ is reached much quicker with rate $-2\Ga(1-\delta)$. Furthermore, for the same reasoning as above, the full model (orange line) agrees for short times better with the local approach, and for larger times better with the non-local approach. Surprisingly, we see that in the full model, the parity takes forbidden values $P>1$. This is connected to the violation of positivity. Although we find that $\tr\rho=1$ is fulfilled for all times $t$, single eigenvalues can become smaller than zero, because positivity is not ensured in the raw model of Eq. \eqref{eq:timelocal}.
\end{example}

To conclude, for cold baths, the different approximations lead to significant differences in the results. Non-local transitions in space play an important role for cold baths and, thus, the local approximations $\LS_\text{loc}$ and $\LC_\text{loc}$ fail. The full model agrees best with the non-local coherent approximation $\LC$. There, differences arise mostly for very short times, where the full model behaves essentially local.

\subsection{Small but finite temperature}

\begin{figure}[ht!]
	\includegraphics[width=0.9\linewidth]{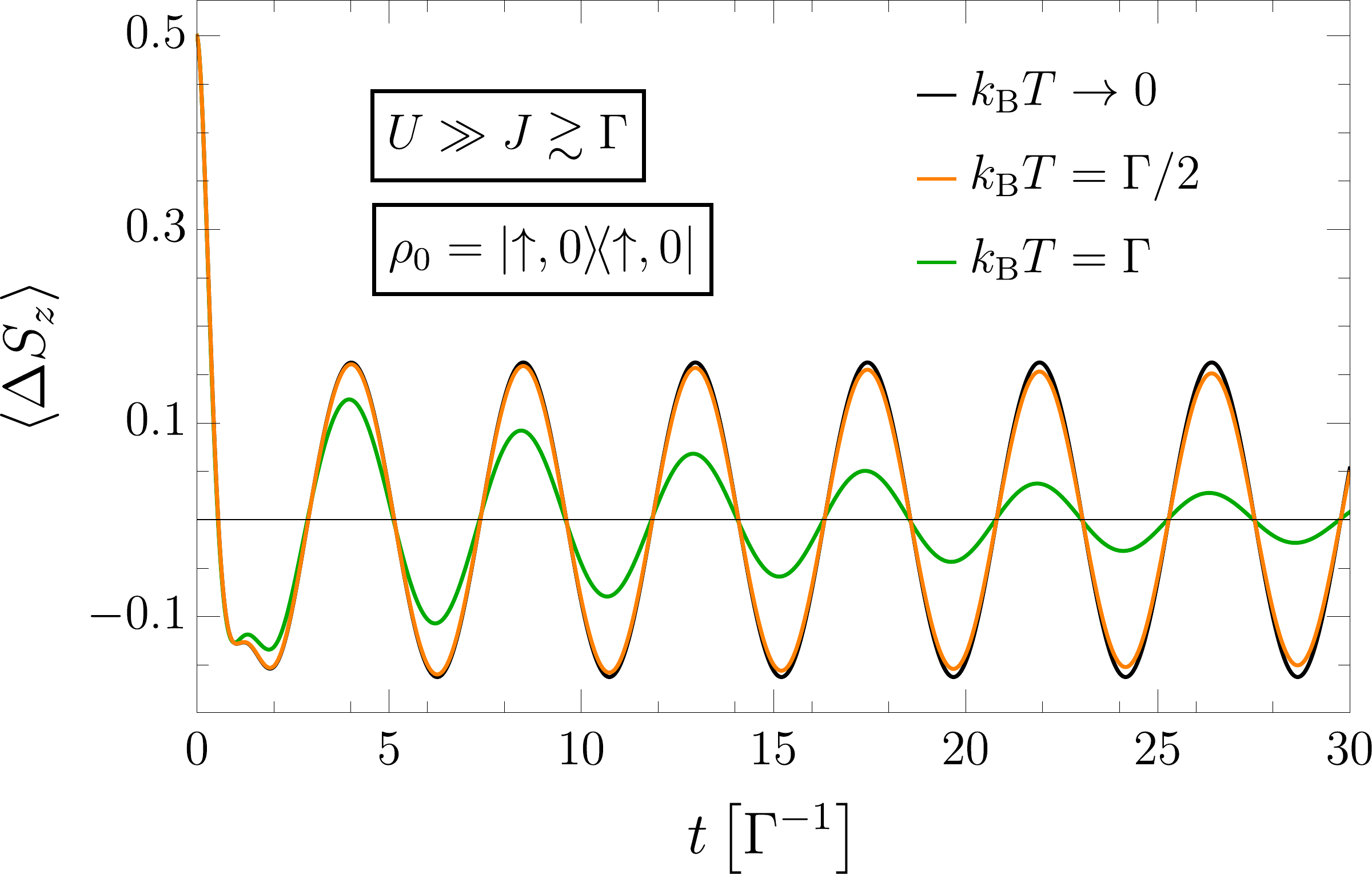}	
  \caption{Spin asymmetry $\Delta S_z$ as a function of time starting from an initial state $\rho_0=\dyad{\su,0}$. We compare the results of the coherent approximation $\LC$ for zero temperature (black line) with finite temperatures $k_\text{B}T= \Ga/2$ (orange line) and $k_\text{B}T=\Ga$ (green line). The spin oscillations decay on a rather long time scale. The remaining parameters are $U=10\Ga$ and $J=2\Ga$.}
  \label{fig:temperature}
\end{figure}

By studying small but finite temperatures, $T>0$, we find that for all methods the $T=0$ result is continuously reached for $T \rightarrow 0$. In Fig. \ref{fig:temperature}, we show as an example $\Delta S_z$ for the coherent approximation $\LC$. There we found that for $T\rightarrow 0$ (black line) coherent spin oscillations persist for all times. By gradually turning on the temperature we find that even upto $k_\text{B}T\approx\Ga/2$ the coherent oscillations survive on a rather long time scale. An appreciable decay can first be seen by choosing $k_\text{B}T\approx\Ga$. Hence, the spin oscillations are robust against temperature and we find an ultra-slow relaxation with an exponentially suppressed decay constant $\sim \Ga \exp\lbrack-U/(2k_\text{B}T)\rbrack$ for $U\gg k_\text{B}T$ which gives zero for either $U\rightarrow \infty$ or $T\rightarrow 0$. Here, we assumed the particle-hole symmetric case $\eps{-}\eps_\text{F}={-}U/2$ and hence $U/2$ is the approximate distance of the excitation energies to the Fermi level.

\section{Conclusions}

We have studied relaxation dynamics for interacting electrons in the Hubbard model when each Hubbard site is individually coupled 
to a thermal bath of electrons.
To describe the coupling between system and bath, we used, on the one hand, a heuristic approach where the anticipated effects of the bath are taken into account in form of \textit{ad-hoc} Lindblad operators that are chosen by pure phenomenological arguments. 
For a microscopic treatment, on the other hand, we employed the diagrammatic real-time technique and derived from it a set of Lindblad operators by assuming 
the Markov limit in leading order perturbation theory in the system-bath coupling.
For one site, we explicitly identified the irreducible diagrams with 
corresponding terms in the Lindblad equation. A subtle but important step was to solve the problem of positivity violation of the reduced density matrix which is well known in this context. However, instead of employing the commonly-used secular approximation, which completely neglects certain types of diagrams, we proposed an alternative and less drastic method named the coherent approximation. It does not neglect any diagrams but minimally modifies them to restore positivity. The coherent approximation becomes exact in the limit $T \rightarrow \infty$  and agrees with the secular approximation for $T=0$, thus, being an improvement to the latter.
Furthermore, we extended the formalism to multiple sites and presented how to obtain both approximations from the diagrams. We found that within the coherent approximation, each process for changing particle number and spin is modeled by one Lindblad operator 
that combines all the available energy channels coherently, while the secular approximation treats them incoherently with one Lindblad operator each. Interestingly, although the couplings to the baths were assumed to be local, the obtained Lindblad operators effectively include non-local transitions. Hence, by assuming local Lindblad operators (as we did in the phenomenological approach), we made an implicit assumption (the local approximation) which turned out to have radical consequences on the dynamics in the small temperature case. Interestingly, the origin of those non-local transitions in space are the time-integrated diagrams, which only on the time scale of the characteristic decay time $\tau_\text{c}$ describe  local transitions. Thus, we find that for very short times the induced transitions are essentially local, while for longer times they include also non-local excitations.

For illustration, we explicitly compared the non-local and local methods for both the secular and coherent approximation by studying the relaxation dynamics of a Hubbard dimer.
For hot baths, $T \rightarrow \infty$, non-local transitions become negligible and the local coherent approximation (which then equals the non-local one) becomes exact. In contrast, the secular approximations (either local or non-local) do miss some qualitative features in the dynamics. For cold baths, $T \rightarrow 0$, we compared the proposed methods for justification with the fully time-dependent first order approach, too. We found that, for early times, the evolution is essentially local. Only for later times, the dissipative dynamics learns about the presence of non-local correlations in the dimer and non-local effects begin to matter strongly. In fact, the local approximations do not even lead to a stationary state that is consistent with the Gibbs state. Furthermore, we found that coherent spin oscillations in the decoherence-free subspace spanned by the triplet and singlet-like states are only well described using the coherent approximation. 

To conclude, for small temperatures, non-local Lindblad operators are crucial, although the physical couplings to the baths are entirely local.  Moreover, the non-local coherent approximation turns out to be generally the most accurate. 

\section*{Acknowledgements}

We gratefully acknowledge funding by the Deutsche Forschungs\-gemeinschaft (DFG, German Research Foundation) -- Project 278162697 -- SFB 1242.

\bibliography{References}

\end{document}